\documentclass{ieeeaccess}
\usepackage{cite}
\usepackage{amsmath,amssymb,amsfonts}
\usepackage{algorithmic}
\usepackage{graphicx}
\usepackage{textcomp}

\usepackage{graphicx} 
\usepackage{listings}
\usepackage{lipsum}

\usepackage[center]{caption}
\newcommand\rev[1]{#1}

\usepackage{threeparttable}
\usepackage{multirow, booktabs, diagbox}
\usepackage{tabularx}
\ifCLASSOPTIONcompsoc
    \usepackage[caption=false, font=normalsize, labelfont=sf, textfont=sf]{subfig}
\else
\usepackage[caption=false, font=footnotesize]{subfig}
\fi

\usepackage{url}

\usepackage{breakurl}
\usepackage[breaklinks]{hyperref}
\def\BibTeX{{\rm B\kern-.05em{\sc i\kern-.025em b}\kern-.08em
    T\kern-.1667em\lower.7ex\hbox{E}\kern-.125emX}}
\begin{document}

\begingroup
\newlength{\xfigwd}
\setlength{\xfigwd}{\textwidth}

\endgroup

\history{Date of publication xxxx 00, 0000, date of current version xxxx 00, 0000.}
\doi{10.1109/ACCESS.2017.DOI}

\title{Defining Security Requirements with the Common Criteria: Applications, Adoptions, and Challenges}
\author{\uppercase{Nan Sun}\authorrefmark{1}\authorrefmark{2}, 
\uppercase{Chang-Tsun Li\authorrefmark{1}, \IEEEmembership{Senior Member, IEEE},
Hin Chan \authorrefmark{3},
Ba Dung Le \authorrefmark{2}\authorrefmark{4},
MD Zahidul Islam \authorrefmark{4},
Leo Yu Zhang \authorrefmark{1}, \IEEEmembership{Member, IEEE},
MD Rafiqul Islam\authorrefmark{4},
and Warren Armstrong}.\authorrefmark{5}}
\address[1]{School of Engineering \& Information Technology, University of New South Wales, Canberra, ACT 2612, Australia}
\address[2]{Cyber Security Cooperative Research Centre, Joondalup, WA 6027, Australia}
\address[3]{Australian Cyber Security Centre, Kingston, ACT 2604, Australia}
\address[4]{School of Computing, Mathematics and Engineering, Charles Sturt University, Wagga Wagga, NSW 2678, Australia}
\address[5]{QuintessenceLabs Pty Ltd, Canberra, ACT 2609, Australia}

\tfootnote{ 
The work is supported by the Cyber Security Research Centre Limited whose activities are partially funded by the Australian Government’s Cooperative Research Centres Programme.}

\markboth
{Sun \headeretal: Defining Security Requirements with the Common Criteria: Applications, Adoptions, and Challenges}
{Sun \headeretal: Defining Security Requirements with the Common Criteria: Applications, Adoptions, and Challenges}

\corresp{Corresponding author: Nan Sun (e-mail: nan.sun@adfa.edu.au).}

\begin{abstract}
Advances \rev{in} emerging Information and Communications Technology (ICT) technologies push the boundaries of what is possible and open up new markets for innovative ICT products and services. \rev{The adoption of ICT products and systems with security properties depends on consumers' confidence and markets' trust in the security functionalities and whether the assurance measures applied to these products meet the inherent security requirements. Such confidence and trust are primarily gained through the rigorous development of security requirements, validation criteria, evaluation, and certification. The Common Criteria for Information Technology Security Evaluation (often referred to as Common Criteria or CC) is an international standard (ISO/IEC 15408) for cyber security.} \rev{Motivated by encouraging the adoption of the CC that is used for ICT security evaluation and certification}, in this paper, we conduct a systematic review of the CC standard and its adoptions. Adoption barriers of the CC are investigated based on the analysis of current trends in cyber security evaluation. In addition, we share the experiences and lessons gained through the recent \textit{Development of Australian Cyber Criteria Assessment (DACCA)} project on the development of the Protection Profile that defines security requirements with the CC. Best practices, challenges, and future directions on defining security requirements for trusted cyber security advancement are presented.
\end{abstract}

\begin{keywords}
Common Criteria, Cyber Security, Protection Profile, Security Standard and Certification, Trusted System
\end{keywords}

\titlepgskip=-15pt

\maketitle
\section{Introduction}
\label{sec:introduction}
\PARstart{S}{tatistics} from the Australian Cyber Security Centre's (ACSC) Annual Cyber Threat Report \cite{ACSA} show a sharp upwards trend in the number of cyber security incidents. Cyber security issues are becoming a day-to-day struggle across both private and public sectors. The ever-increasing number of cyber attacks and security incidents continues to deepen concerns over data breaches, physical system damage, economic loss, reputation harm, and even compromise of national security \cite{sun2018data}. With such acute concerns, the development of security requirements for Information and Communications Technology (ICT) products is of paramount importance. With a robust security infrastructure in place, hacking and other forms of cyber attacks can be prevented and mitigated to some extent \cite{matheu2020survey}. However, security vulnerabilities often slip into ICT products during the development and implementation stages. With the more widespread use of ICT products, it is imperative to have in place a rigorous security process to ensure the products are secure during the design and development process, validate their security performance, and promote the enforcement of protection policies.

\rev{In recent years, the interest in trusted systems that enforce a given set of attributes to a stated degree of assurance has reemerged \cite{stallings2012computer}. Enhancing the trust and confidence users have in ICT products is of great significance in the area of cyber security, which can be gained through setting security standards and using independent assessment against the standards \cite{popek1979encryption}. The Common Criteria for Information Technology Security Evaluation (often referred to as Common Criteria or CC) is an international standard (ISO/IEC 15408) for achieving cyber security certification. It is inherited from ICT security assurance through a rigorous verification process, which is conducted on a case-by-case basis \cite{CC}. With the strict, standardized and repeatable 
methodology, the CC provides assurance for implementing, evaluating and operating a security product at the level that is commensurate with the operational environments.}

Under the CC, vendors list the intended security functional requirements (SFRs) within a Security Target (ST) \cite{CC}. Since new products are constantly being developed and every product is designed and developed differently, Protection Profiles (PPs) have been created for common products, such as databases, operating systems, and smart cards \cite{CCpart2}. Generally, a PP defines a set of security requirements and objectives for a specific category of products or systems. In addition, the PP can serve as a benchmark in terms of product security. Once validated by competent and licensed laboratories \cite{laboratories}, a certificate is issued by the certification authority and recognized by CC signatory countries.

Since the CC standard emerged in the 1990s, there have been 17 Certificate Authorizing Participants (including Australia) and 14 Certificate Consuming Participants signed up to the Common Criteria Recognition Arrangement (CCRA) \cite{CC2}. Suppose an ICT security product is successfully evaluated. In that case, the product will be certified by the certification authority of a CCRA signatory country and listed on the Certified Products List at the CC Portal \cite{CC3}. The certification helps consumers determine whether the products meet their security requirements, which also boosts the competitiveness of the products by comparing them with similar products on the market.

\textbf{Contributions of this paper:}
\rev{This paper aims to provide an overview of current cyber security efforts to develop CCs internationally and in Australia, in order to encourage the adoption of the CC by the wider community. We firstly introduce the CC methodology and contemporary applications of the CC.} Besides, we investigate the current adoption of the CC. By comparing the CC with the state-of-the-art security standards, we explain the significance and demonstrate the impact of the CC. Specifically, through the collaboration with the Australian Certification Authority of the Australian Cyber Security Centre, QuintessenceLabs and cyber security researchers in academia, the lessons and best practices relevant to defining security requirements with the PP  development are presented in this paper. The target audiences of the paper are researchers in academia, security policy-makers, industrial practitioners, and end-users in public and private sectors who are interested in the specification, development, evaluation, certification, procurement, and operation of ICT products with security properties. In summary, the contributions of this work are as follows.

\begin{itemize}
    \item A systematic review of the CC and its applications are demonstrated based on literature review combined with practical experience gained through the recent \textit{Development of Australian Cyber Criteria Assessment (DACCA)} project \footnote{https://cybersecuritycrc.org.au/development-australian-cyber-criteria-assessment}. 
    \item An in-depth and comprehensive analysis of current trends in the CC adoption is carried out. Based on the identified challenges of the CC adoption, the adoption barriers of the CC internationally and nationally are analyzed. 
    \item Practices, recommendations and future directions derived from the analysis and solutions to address identified challenges in defining security requirements with the CC are presented.    
\end{itemize}

\rev{
\textbf{Comparison with related works:}
There are limited review works that explore security requirements for ICT products and services. The latest work in \cite{matheu2020survey} provided an overview of cyber security certification for the Internet of Things (IoT). In \cite{matheu2020survey}, Matheu et al. analyzed the various cyber security certification schemes and the potential challenges in applying them to the IoT ecosystem. They also studied current efforts in risk assessment and testing processes. The work \cite{matheu2020survey} made significant contributions to the deployment of an IoT cyber security certification framework. However, this work \cite{matheu2020survey} focused specifically on IoT products and covered a broad range of certification standards without an in-depth discussion of the CC methodology and adoption. From the perspective of security testing and risk assessment in the IoT, previous works \cite{bures2018internet} \cite{dias2018brief} \cite{kuzminykh2018analysis} reviewed the key building blocks for the cyber security certification process. Besides security certification in IoT, Leszczyna et al.  \cite{leszczyna2018cybersecurity} conducted a comprehensive survey on smart grid standards that deal with cyber security issues and provided valuable insights into security-related standards. The work in \cite{leszczyna2018cybersecurity} covered 36 cyber security-related and 12 privacy-related standards on smart grids. However, similar to \cite{matheu2020survey}, insightful analysis on the CC is lacking. In addition, Kara et al. \cite{kara2012review} reviewed the CC in a specific field, which stressed CC's applications in secure software development.  When it comes to the significance of the CC, Matheu et al. \cite{matheu2020survey} acknowledged the CC as the most widely deployed and adopted cyber security certification standard in the field of IoT. Furthermore, Houmb et al. \cite{houmb2010eliciting} proposed a CC-driven security requirements elicitation and tracing approach, which demonstrates the capability of the CC on providing security expertise, knowledge, and guidelines for building secure systems. Albeit the fact that Russia is neither a Certificate Authorizing Participant nor a  Certificate Consuming Participant of the CC, the history, structure, and features of the CC used in the Russian scheme are presented in \cite{barabanov2015modern} and \cite{barabanov2014russian}, which manifest the importance of the CC. In addition, China is not a CCRA signatory country but has the adoption of the CC called GB/T 18336 \cite{hu2019summary}. To the best of our knowledge, this paper is the first comprehensive survey on the CC for ICT Security Evaluation with regard to its applications, adoptions and related challenges.  }

\textbf{Roadmap:}
The rest of the paper is organized as follows. We start with introducing the methodology and applications of the CC towards security assurance for ICT product evaluation in Section 2. Section 3 presents our comprehensive literature review with respect to the CC adoption trend, adoption barriers, and its impact. Section 4 discusses the challenges in PP development and our experience gained from the development of a Protection Profile that defines the security requirements for encryption key management appliances. In addition, the best practices and future directions that support CC approaches are shared based on the identified challenges. Section 5 concludes this review paper.

\section{Common criteria methodology}
This section aims to introduce the CC methodology and its applications for evaluating ICT products. The analysis is based on Protection Profiles and Certified Products listed on the CC portal, research papers and technical reports from various organizations including governments, vendors and other CC participants. In addition, potential categories of CC applications on emerging technologies are proposed.

\subsection{Driving security assurance through Security Targets and Protection Profiles}
The CC is the driving force for the widest available mutual recognition of secure ICT products. In this subsection, we demonstrate how to drive security assurance through Security Targets and Protection Profiles with the CC standards. Firstly, definitions of key CC concepts are identified and specified. Secondly, we introduce the methodology of the CC by presenting the rationales and relationships among the core building blocks in the CC. Thirdly, we emphasize that the CC is risk-based by illustrating how the CC works to reduce and minimize risks and threats, so as to raise users' confidence in the security performance of ICT products.

\subsubsection{Background and Definitions}

The CC was developed to certify that products and systems meet pre-defined security requirements \cite{herrmann2002using}. Through a set of specifications and guidelines designed to evaluate ICT products and systems, the products that have undergone successful testing and evaluation are awarded the CC certification \cite{stallings2012computer}.

\textbf{The history of the CC:} In 1994, the CC was developed by the governments of the US, Canada, Germany, France, the UK, and the Netherlands \cite{herrmann2002using}. The CC is the unified standards of the Canadian Trusted Computer Product Evaluation Criteria (CTCPEC), the United States Trusted Computer System Evaluation Criteria (TCSEC), and the European Information Technology Security Evaluation Criteria (ITSEC) \cite{historyCC2}. The consolidation of these security standards helps to avoid the repetitive work on the evaluation of similar products and systems of the same type and also addresses the application in prevailing international markets. By March 2021, the Common Criteria Recognition Arrangement (CCRA) consists of 31 governmental organizations, including 17 certificate authorizing nations and 14 certificate consuming nations \cite{aboutCC}. The CCRA aims to conduct evaluations to high and consistent standards, improve the availability of Certified Products, and eliminate the duplication and improve the efficiency of evaluations and processes \cite{aboutCC}. The CCRA maintains the CC portal's Certified Products List (CPL) \cite{CC3} that lists all CC Certified Products completed by all certificate authorizing nations.

\begin{table}
\centering
\caption{The classes of Security Assurance Requirements}
\label{SAR}
\begin{tabular}{cc}
\hline
Class&Class name\\
\hline
APE&Protection Profile Evaluation\\
ASE&Security Target Evaluation\\
ADV&Development\\
AGD&Guidance Documents\\
ALC&Life-Cycle Support\\
ATE&Tests\\
AVA&Vulnerability Assessment\\
ACO&Composition\\
\hline
\end{tabular}
\end{table}

\begin{table}
\centering
\caption{The classes of Security Functional Requirements}
\label{SFR}
\begin{tabular}{cc}
\hline
Class&Class name\\
\hline
FAU&Security Audit\\
FCO&Communication\\
FCS&Cryptographic Support\\
FDP&User Data Protection\\
FIA&Identification and Authentication\\
FMT&Security Management\\
FPR&Privacy\\
FPT&Protection of the TOE Security Functionality\\
FRU&Resource Utilization\\
FTA&TOE Access\\
FTP&Trusted Path/Channels\\
\hline
\end{tabular}
\end{table}

\textbf{Key CC concepts:} The part of the product or system that is the subject of the evaluation is called the Target of Evaluation (TOE). To define a standard set of security requirements for a particular class of related products, the Protection Profile (PP) is usually developed by a user or a user group \cite{herrmann2002using}. A PP serves as a reusable template of security requirements to support the definition of functional standards, and also as a guide for formulating product development or procurement specifications. A PP is an implementation-independent set of security requirements for a particular technology that enables repeatable evaluations. To enhance the consistency of testing, some PPs are augmented with the specification of testing activities. Depending on the TOE, multiple profiles can be used at once based on the particular technology that the TOE is to be certified. If a vendor has an ICT product that they would like to be evaluated and certified under CC standards, they must complete a Security Target (ST) description. The ST is the document provided by the vendor to identify the security features of the TOE \cite{CC}. In addition, the ST includes the evaluation of any potential security risks by defining the security functional and assurance measures that the TOE should offer to meet CC requirements.

As shown in Table \ref{SAR}, eight categories of Security Assurance Requirements (SARs) are identified by the CC to be used as the basis for gaining confidence that claimed security measures are implemented correctly. The CC defines eleven categories of Security Functional Requirements (SFRs) in relation to desirable security functionalities to provide a standard way of expressing the requirements for a TOE, as summarized in Table \ref{SFR}. The Evaluation Assurance Levels (EAL) define how the product is tested and how thoroughly the product is evaluated. The EAL levels are scaled from EAL1 (the lowest) to EAL7 (the highest) \cite{CC}. It should be noted that the EAL number does not measure the security of the product but states at what level the product or system was tested. A higher EAL level reflects added assurance requirements that must be met to achieve the CC certification \cite{CCpart3}. Although the product or system that will be certified must fulfill the exact \textit{assurance} requirements to achieve a specific EAL level, they do not have to fulfill the exact \textit{functional} requirements (i.e., security features). Therefore, the product or system with a higher EAL level does not necessarily mean more secure in the particular application than the one with a lower EAL level. If two products contain the same and necessary \textit{security features} in the ST, then a higher EAL level indicates the product is more secure.

\textbf{How are products tested:} For evaluation against a PP that specifies testing activities, the vendor should complete a self-assessment on compliance with the PP. For EAL-based evaluation, the vendor's testing would serve as an input to the evaluator's testing \cite{CC}. The tests are carried out under laboratory conditions to validate the security features of the product and to evaluate how the product satisfies the requirements listed in the ST \cite{laboratories}. If the validation and evaluation are successful, the product will be awarded a CC certificate and listed on the CC portal \cite{CC_portal}. From the consumer's point of view, the CC certification ensures consumers that they can trust the products they are investing in conformance to the vendor's claims and can offer reliable security protection for their operational environment. For the vendors, the CC certification boosts the competitiveness of their products when the consumers compare similar products on the market. One of the advantages of using the CC is that products can be evaluated once and sold in multiple nations. The CCRA ensures that the same criteria and testing methodology are applied to the products against the same standards in different accredited laboratories, regardless of their geographic location or national affiliation. For governments, besides supporting procurement, the CC certification also increases the transparency of ICT products' security features, facilitating the supervision and surveillance of the market.   

\begin{figure}
  \centering
  \includegraphics[width=\linewidth]{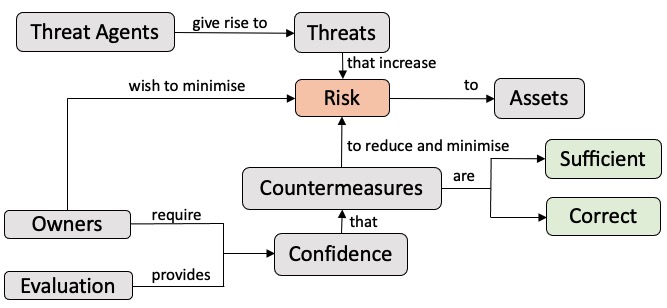}
  \caption{The Common Criteria is risk-based}
  \label{risk}
 
\end{figure}
\textbf{The CC is risk-based:}
The CC process is helpful as a guide for the development, evaluation and procurement of ICT products with security functionality \cite{CCpart3}. Typically, from the perspective of risk control and management, the CC is risk-based, as illustrated in Figure \ref{risk}. On the one hand, under the CC methodology, security is concerned with protecting assets that refer to entities that the owners of a system places value upon, including hardware, software, data, and transmission link. On the basis of the impact of the threats on the assets and the likelihood of the threats being exploited, threats increase the risks to the assets. The owners impose ICT and non-ICT countermeasures that seek to reduce and minimize the risk to assets. On the other hand, the CC evaluation provides the confidence to achieve the protection goals of ICT security with confidentiality, integrity, availability, authenticity, and non-repudiation \cite{von2013information}, which is also needed by vendors and purchasers. Sufficient and correct countermeasures, which are achieved by conforming to CC requirements, will minimize the risks to the assets. By evaluating ICT security assessments in line with CC, trust under risk can be achieved.

\subsubsection{Methodology of the Common Criteria} 

\begin{figure*}
  \centering
  \includegraphics[width=0.7\linewidth]{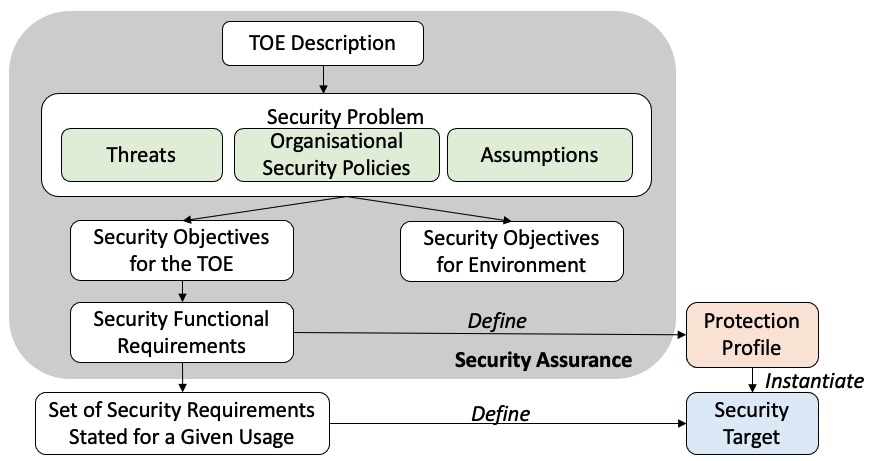}
  \caption{Methodology of the Common Criteria}
  \label{methodology}
\end{figure*}
\rev{We explain the CC methodology by introducing core building blocks of the CC, including TOE description, security problems, security objectives, and security requirements. The relationships between these core building blocks are depicted in Figure \ref{methodology}. Generally, a Protection Profile (PP) defines the security requirements of a technology type while a Security Target (ST) describes how the TOE meets the defined requirements in the CC. The TOE physical environment, security problems, security objectives, security requirements and the purpose of the TOE, which is relevant to the product type and the intended use, are included in the PP \cite{matheu2020survey}. The goal of the CC methodology is to achieve an internationally recognized evaluation benchmark for ICT security. As shown in Figure \ref{methodology}, a PP is a CC requirement specification for a specific technology. In the context of a dedicated use-case, a set of security requirements stated for a given usage are extracted to instantiate an ST and determine the TOE \cite{hennebert2020first}.}

\textbf{TOE description:} To give end users a general understanding of what the TOE can do, the way it can be used, and whether it meets their security needs \cite{CC} is the first step in the CC methodology.

\textbf{Security problems:} In line with the description of the TOE, the security problems will be defined in three aspects: threats, organizational security policies, and assumptions \cite{CC}. Firstly, a threat consists of adversarial actions performed by a threat agent on an asset. The actions affect one or more properties of the asset from which that asset derives its value. One example of a threat is ``An attacker may attempt to intercept communications from the TOE". Secondly, organizational security policies are security rules, procedures, or guidelines imposed by an actual or hypothetical organization in the operational environment, which can be applied to the TOE and/or its operational environment. An example of organizational security policies on the auditing activities is ``Audit logs will be archived every seven days". Lastly, assumptions can be made about the operational environment, including the physical setting (e.g., ``the TOE is located in a secure area, which will prevent unauthorized physical access"), personnel skills and behaviour (e.g., ``operators of the TOE are appropriately trained"), and connectivity (e.g., ``the TOE will not be connected to an insecure network").

\textbf{Security objectives:} The security objectives are the intended solution expressed as a concise and abstract response to the security problems. The security objectives serve three purposes:  (1) provide a high level solution to the security problems; (2) divide the solution into those objectives to be satisfied by the TOE, and those objectives to be satisfied by the environment; (3) demonstrate that these part-wise solutions can form a complete solution to the security problem. Security objectives for the TOE consist of the objectives that the TOE should achieve to solve security problems. For example, a high-level objective such as ``the TOE shall keep confidential the content of all files transmitted between it and a server" addresses the security problems caused by unauthorized monitoring of network traffic. The security objectives for the operational environment include statements that describe the goals that the environment should achieve. For example, ``the operational environment shall ensure that all human TOE users receive appropriate training before working with the TOE" provides the procedural measures to assist the TOE in providing its security functionalities correctly from the perspective of the operational environment. 

\textbf{Security Functional Requirements (SFRs):} The SFRs are a translation of the security objectives for the TOE \cite{CCpart2}. The SFRs provide a more detailed and complete translation to make sure that the security objectives can be completely addressed. Specifically, the SFRs are independent of any specific technical implementation. Hence, the CC requires the translation from security objectives to the SFRs to be conducted with standardized language. The advantages and reasons for the standardized language requirements are twofold. One is to provide an exact description instead of natural language of what is the functionality to be evaluated. Another advantage is to allow comparison among products of the same class that will be evaluated. The standardized language enforces the use of the same terminology and concepts that contributes to the easy comparison. The CC supports the standardized translation by providing predefined security functional requirements, operations, and dependencies. It is worth mentioning that the SFRs are only for the TOE. The operational environment is not evaluated, and therefore a description aimed at the evaluation of the operational environment is not required \cite{CCpart2}. The parts of the operational environment may be evaluated in another independent evaluation. For instance, an operating system as a TOE may require a firewall to be present in its operational environment. The current evaluation focuses on the operating system TOE only, and another evaluation can subsequently be applied to the firewall.

\begin{table*}
\centering
\caption{\rev{The summary of Common Criteria applications, including current categories and future developments}}
\label{categories}
\begin{tabular}{l|l|l|l} 
\hline
                                                                                & Category                                                                                              & Example products                                                                                                                                 & Research Efforts                                                                                                                                             \\ 
\hline
\hline
\multirow{15}{*}{\begin{tabular}[c]{@{}l@{}}Existing~\\Categories\end{tabular}} & Access control devices and systems                                                                    & \begin{tabular}[c]{@{}l@{}}Single Sign On (SSO) and identiy managers~\end{tabular}                                       & \cite{singh2010formal}                                                                                                                      \\ 

\cline{2-4}
                                                                                & Biometric systems and devices                                                                         & Palm vein biometric systems                                                                                                                      & \cite{tekampe2011d6}                                                                                                                        \\ 
\cline{2-4}
                                                                                & Boundary protection devices and systems                                                               & Security gateways                                                                                                               & \cite{xu2019co}                                                                                                                             \\ 
\cline{2-4}
                                                                                & Data protection                                                                                       & \begin{tabular}[c]{@{}l@{}}Data transport systems\end{tabular}                           & \cite{khan2000characterising} \cite{kindt2016privacy} \cite{meints2008biometric}                          \\ 
\cline{2-4}
                                                                                & Databases                                                                                             & \begin{tabular}[c]{@{}l@{}}Distributed databases and graph databases\end{tabular}                                                               & \cite{mohammad2010index}                                                                                                                    \\ 
\cline{2-4}
                                                                                & Detection devices and systems                                                                         & \begin{tabular}[c]{@{}l@{}}Intrusion detection systems scanners and analyzers\end{tabular}                                                      & \cite{lee2013security}                                                                                                                      \\ 
\cline{2-4}
                                                                                & \begin{tabular}[c]{@{}l@{}}ICs, smart cards and smart card-related devices\\ and systems\end{tabular} & \begin{tabular}[c]{@{}l@{}}JavaCard, electronic residence permits, and electronic passports\end{tabular}                                        & \cite{narasamdya2009certification}                                                                                                          \\ 
\cline{2-4}
                                                                                & Key management systems                                                                                & {Public key infrastructures}                                                                                    & \cite{DACCA_report2}                                                                                                                       \\ 
\cline{2-4}
                                                                                & Mobility                                                                                              & Smartphones, tablets, and laptops                                                                                                                & \cite{lee2015security}                                                                                                                      \\ 
\cline{2-4}
                                                                                & Multi-function devices                                                                                & Hardcopy devices                                                                                                                                 & \cite{lee2015study}                                                                                                                         \\ 
\cline{2-4}
                                                                                & Network and network-related devices and systems                                                       & Hubs, switches, and routers                                                                                                                      & \cite{smith2007trends}                                                                                                                      \\ 
\cline{2-4}
                                                                                & Operating systems                                                                                     & Systems and servers                                                                                                                         & \cite{toll2008tooling}                                                                                                                      \\ 
\cline{2-4}
                                                                                & Products for digital signatures                                                                       & Signature creation devices                                                                                                                                         & \cite{langenstein2000use}                                                                                                                   \\ 
\cline{2-4}
                                                                                & Trusted computing~                                                                                    & Trusted platform modules                                                                                                                          & \cite{lohr2009modeling}                                                                                                                     \\ 
\cline{2-4}
                                                                                &Other devices and systems                                                                             & \begin{tabular}[c]{@{}l@{}}Web browsers, voting machines, and smart TVs\end{tabular}                                                            & \cite{kang2017obtain}                                                                                                                       \\ 
\hline
\multirow{5}{*}{\begin{tabular}[c]{@{}l@{}}Potential\\Categories\end{tabular}}  & Blockchain mechanisms and systems                                                                     & Blockchian platforms\textcolor[rgb]{0.125,0.129,0.141}{~}                                                                                        & \cite{matsuo2017formal}                                                                                                                     \\ 
\cline{2-4}
                                                                                & Quantum computing                                                                                     & Quantum computers                                                                                                                                 & \cite{cadzow2018analysis}                                                                                                                   \\ 
\cline{2-4}
                                                                                & Privacy-preserving authentication                                                                     & \begin{tabular}[c]{@{}l@{}}{Privacy-preserving biometric authentication}\end{tabular} & \cite{beckers2012foundation}                                                                                                                \\ 
\cline{2-4}
                                                                                & Artificial intelligence systems                                                                       & Facial Detection and Recognition systems                                                                                                          & \cite{AIproposal}                                                                                                                          \\ 
\cline{2-4}
                                                                                & Internet of Things applications                                                                       & Smart city and smart homes                                                                                                                       & \cite{kang2017obtain} \cite{kim2017design} \cite{matheu2020survey} \cite{suciu2019lego}  \\
\hline
\end{tabular}
\end{table*}

\textbf{Security Assurance Requirements (SARs):} Different from the SFRs, SARs describe the measures taken during the process of development and evaluation of ICT products to assure compliance with the security functionalities \cite{CCpart3}. The pre-packaged set of SARs is defined in the CC as summarized in Table \ref{SAR} and specified from EAL1 to EAL7. For certain PPs, the EAL specification is optional, which indicates the PP can specify a customized set of assurance components (i.e., no EAL). The recent trend on the PP development includes assurance activities for each SFR by specifying detailed actions in the supporting documents for the evaluator to perform \cite{CC_portal}. It is noted that the TOE always makes assumptions about the operational environment. Security objectives for the TOE do not trace back to assumptions, and they are not evaluated but need to be understood and upheld.

\subsection{Common Criteria applications: current categories and future developments}
In this subsection, we provide the analysis of CC applications by reviewing current efforts. In addition, we propose future developments for the CC applications based on emerging technologies and compliance requirements with emerging privacy legislation. \rev{Table \ref{categories} summarizes existing and potential CC applications with the illustration of example products and references of research efforts.} 

\subsubsection{Existing Categories}
\label{current_categories}

We firstly review and summarize the existing Certified Products and applications under the CC by category. The Certified Products and Protection Profiles endorsed on the CC Portal \cite{CC_portal} by December 2021, research papers, and technical reports on the CC are included in the review.
\rev{

\textbf{Access control devices and systems:} Functioning within the framework of the security system, access control devices and systems ensure only authorized persons can access the system \cite{bowers2013access}. Usually, an access control system is a software-based application that provides an interface for authorized users to pass through an interface integrated into the system. The access control devices are the physical hardware that an access control system requires to enforce the functional rules. Specifically, Singh et al. \cite{singh2010formal} proposed a formal security policy model for implementing the insider threat protection security solution for the network computing environment in line with CC. Under the category of access control devices and systems, the CC portal \cite{CC_portal} contains PPs for the evaluation of Single Sign On (SSO) and enterprise management access control. The archived PP list on the CC portal \cite{CC_portal} lists the 11 PPs for firewalls, intrusion detection systems and the US Government Authorization server for basic robustness environments, which are for reference only and are not to be used as a basis for new evaluations. In addition, 25 products are certified under the access control devices and systems category. 
%


   

\textbf{Biometric systems and devices:} Biometrics is one of the most robust and reliable approaches for human identification in the physical and cyber spaces \cite{hill2015wearables}. Tremendous advances in sensor technologies and data processing techniques lead to the strengthening of traditional biometric technologies (e.g., fingerprint, face, voice, iris, etc.) and the emergence of new technologies (e.g., DNA analysis, biometric payment cards, etc.). However, vulnerabilities and threats will inevitably occur, highlighting the significance of evaluating the security of biometric systems and devices \cite{alaswad2014vulnerabilities}. The PPs for fingerprint spoof detection based on organizational security policies and biometric verification mechanisms are listed on the CC portal \cite{CC_portal}. So far, there are no Certified Products under this category. Tekampe et al. \cite{tekampe2011d6} offered guidance for the evaluators of biometric system, vendors, and certifiers of biometric systems and devices according to the CC for security evaluation, which is a valuable input for further standardization activities.

\textbf{Boundary protection devices and systems:} Boundary protection monitors and controls communications at the external boundary of the system to prevent and detect malicious and other unauthorized communication \cite{kissel2011glossary}. Boundary protection can be achieved through firewalls, routers, gateways, proxies, and encrypted tunnels \cite{adler2019remote}\cite{xu2019co}. Practical design, installation, configuration, and maintenance of the boundary protection devices and systems are critical tasks in providing effective cyber security. In summary, there are a total of 46 Certified Products and 42 PPs listed on the CC portal in this domain, including several personal firewalls. One example of PPs in this category is the collaborative Protection Profile (cPP) Module for Stateful Traffic Filter Firewalls. Typically, a cPP is a PP that has been created through a collaborative process consisting of vendors, test laboratories, CCRA nations, and academia to define requirements and testing methodology through industry engagement.

 \textbf{Data protection:} Data protection refers to the rules, safeguards, and practices put in place to protect data and ensure that users remain in control of the data \cite{bygrave2002data}. As the extent and potential value of data increases, the data protection regulation is significant for users to protect the privacy of users \cite{wachter2019data}. There are 28 current data protection PPs available on the CC portal, which cover cryptographic modules, encrypted storage device and cryptographic protocols. There is also a cPP for full drive encryption. There are 66 Certified Products under the category of data protection. In addition, there are a number of research efforts on the data protection for the CC development \cite{khan2000characterising} \cite{kindt2016privacy} \cite{meints2008biometric}. With the aid of functional requirements defined in the CC for data protection, Khan et al. \cite{khan2000characterising} characterized user data protection of software components to boost the confidence and trust in component technologies. Furthermore, privacy and data protection issues of biometric applications are always a recurring question when applying existing data protection legal frameworks to respond to the new threats under the fundamental rights to respect for privacy and data protection. Kindt \cite{kindt2016privacy} systematically analyzed the privacy and data protection issues of biometric applications and summarized the key requirements according to the CC for Information Technology Security Evaluation. To develop and run the biometric systems in compliance with European data protection legislation, Meints et al. \cite{meints2008biometric} investigated the most relevant data protection principles in the field of biometric systems. 

\textbf{Databases:} A database is the organized collection of structured data stored in the computer system \cite{kim1979relational}. Typically, a database is controlled by a database management system \cite{mohammad2010index}. A relational database is the most common type of database systems, including Structured Query Language (SQL) server, Oracle Database, MySQL, etc. In addition, there are other types of databases available in the market nowadays, including NoSQL databases \cite{davoudian2018survey}, distributed databases \cite{ozsu1996distributed}, graph databases \cite{angles2008survey}, cloud databases \cite{deka2013survey}, centralized databases and commercial databases. Currently, there are 11 database management systems related PPs and one cPP available on the CC portal and 14 Certified Products, including the industry-leading advanced  databases from IBM, Microsoft, Oracle, and HUAWEI.

\textbf{Detection devices and systems:} Intrusion detection and prevention system protects users' ICT systems and applications by identifying suspicious activity and behavior \cite{lee2013security}. Detection devices and systems usually operate by monitoring and analyzing network traffic and providing proactive and preventive measures to ensure the security of the machines on which they are deployed \cite{liao2013intrusion}. There are 17 expired and archived PPs on intrusion detection systems scanners and analyzers, while there is no current PP on the CC portal in this stage. Moreover, nine products that do not conform to any PPs under the category of detection devices and systems are certified through EAL-based evaluation.

\textbf{ICs, smart cards and smart card-related devices and systems:}  The category which sees most comprehensive application of the CC is Integrated Circuits (ICs), smart cards and smart card related devices and systems. ICs and smart cards are physical electronic authorization devices used to access a resource \cite{shelfer2002smart}. Usually, it is a card with an embedded integrated circuit, which may require physical contact or can be contactless \cite{narasamdya2009certification}. There are 84 current PPs covering JavaCard, electronic residence permits, electronic passports, Machine Readable Travel Documents (MRTDs), security module cards, and health cards. There is a cPP for the dedicated security component. There are 571 Certified Products under this category.

 \textbf{Key management systems:} Key management systems refers to the management of cryptographic keys in cryptosystems \cite{ghosal2019key} \cite{hegland2006survey} \cite{xiao2007survey} . Key management systems and appliances are designed to centrally manage enterprise digital keys and certificates for enterprise applications, users and devices throughout their lifecycle. A key management system handles key generation, distribution, usage, automated rotation, renewal and revocation. Therefore, successful key management is critical to the security of a cryptosystem. Limited PPs and Certified Products are included in the CC portal. However, a secure, usable, unified and centralized suite of complementary requirements for key management systems is expected by vendors, purchasers and research community \cite{stieber2010enterprise}. Based on our experience gained through the \textit{DACCA} project \cite{DACCA_report2}, we will review and summarize the lessons learned from the practical PP development on encryption key management components in Section 4.

\textbf{Mobility:} The mobility category covers the evaluations of mobile device fundamentals and mobile device management. The mobile device provides essential services, such as cryptographic services, key storage services, and data-at-rest protection to support the operation of applications on the device securely \cite{gao2015study}. Furthermore, a mobile device management is the administration of mobile devices, including smartphones \cite{lee2015security}, tablets, and laptops \cite{hayes2020effective}. A total of 27 Certified Products are evaluated under the CC standard, including the mobile devices from Samsung, Google, Apple, Blackberry and other mobile devices representatives.     

\textbf{Multi-function devices:} A Multi-Function Device (MFD) refers to an equipment that can print, copy and scan\cite{lee2015study}. Threatened by vulnerabilities in relation to network connections, MFD devices may be laden with security vulnerabilities and may fall victim to security incidents like data exposure and eavesdropping \cite{lee2015study}. There are 229 MFD certified devices listed on the CC portal and five PPs for hardcopy devices available as part of certification processes according to the CC. 

\textbf{Network and network-related devices and systems:} Network devices are physical devices that are needed for communication and interaction between hardware on the computer network \cite{vieira2003survey}. Common network devices include hubs, switches, routers, gateways, bridges, modems, repeaters, and wireless access points \cite{smith2007trends}. There are 232 Certified Products and 13 current PPs under the network and network-related devices category, there is one cPP for the network devices \cite{NDcPP} that was developed by the networking international Technical Community (iTC) and was updated over time. 

\textbf{Operating systems:} An operating system is the system software that manages computer hardware, software, and contributes common services for computer programs \cite{toll2008tooling}. This category includes a considerable number of PPs and Certified Products (e.g., MacOS Catalina 10.15, Windows 10 and Windows Server 2019 Version 1809, HongMeng V1.2, etc.) for operating systems in networked environments. 
 
\textbf{Products for digital signatures:} The digital signature is a mathematical algorithm for validating the authenticity and integrity of digital messages or documents \cite{paul2017digital}.  Products for digital signatures are one of the application categories of the CC evaluation \cite{langenstein2000use}. There are 53 Certified Products available on the CC portal, such as DocuSign. Within this category, there are also 23 PPs for cryptographic modules and digital signature creation devices.

\textbf{Trusted computing:} Trusted computing refers to technologies developed for resolving network security problems by enhancing hardware and modifying associated software components. The computer industry has accommodated the idea of trusted computing that is designed and promoted by Trusted Computing Group (TCG) in various ways. The TCG published the Trusted Platform Module (TPM) specification and a corresponding PP, which represents efforts to develop formal criteria for evaluating its security \cite{trustedPP}. Although some people argue that trusted computing is unlikely to become a complete remedy for security problems \cite{oppliger2005does}, there has been a sharp rise in the number of PPs and Certified Products under the category of trusted computing in recent years. Löhr et al. \cite{lohr2009modeling} demonstrated how to advance PP development for trusted computing technology with exemplar projects.

\textbf{Other devices and systems:} This category contains Certified Products and PPs for everything else which do not fit in the afore-mentioned categories, including web browsers, voting machines, smart TVs, etc. Research efforts, for example, the discussion on how to obtain CC certification for smart TV, are found in the literature \cite{kang2017obtain}.

}
\subsubsection{Potential Categories}

\label{potential_categories}
The increasing adoption of emerging technologies motivates the study of potential categories of the CC. We list below the emerging technologies and compliance requirements with privacy legislation that may be included as new CC categories in the future. 

\textbf{Blockchain mechanisms and systems:} Blockchain offers innovative and integrated approaches to ensure information storage and transactions executed in an open environment are easily verifiable and auditable to all participants \cite{kolb2020core}. Blockchain is considered a ground-breaking technology for cryptography and cyber security, with application in many areas including  cryptocurrency systems, smart contracts, and smart grids over IoT devices. \cite{sharma2020blockchain}. However, the security and privacy concerns attributed to the blockchain technology should not be ignored when deploying blockchain in different applications. Furthermore, the maturity of the blockchain technology and relevant protocols are not sufficient to subside security concerns without subjecting blockchain-enabled IT products to standardized security evaluation and validation\cite{zhang2019security}. For example, Matsuo \cite{matsuo2017formal} pointed out that the application logic layer of the blockchain technology, that contains the scripting language for the financial transaction does not yet have a standard to provide security analysis. 

\textbf{Quantum computing:} The emerging technology of quantum computing encodes information in qubits as a non-classical approach, enabling computing to be conducted $2^{n}$ times faster than classical computing \cite{gill2020quantum}. There are considerable speculations from industry and academia about the impact of quantum computing on cyber security \cite{cadzow2018analysis}. In light of the potential power of quantum computers at non-trivial scale, it will be important to study and explore the incorporation of quantum-resistant algorithms or alternative approaches (such as quantum key distribution) into CC requirements around network security.  These requirements will require modification to ensure Certified Products can continue to meet security guarantees around confidentiality and integrity of their data, particularly if transmitted over an untrusted network between trusted endpoints.   In addition, the investigation into zero knowledge proofs' reliance on post-quantum hardness assumptions for security should be considered in the CC development. 

\textbf{Privacy-preserving authentication:} In the last few years, many privacy-preserving authentication methods have been proposed to make authentication technologies reliable and secure \cite{karger2010lessons} \cite{wang2016challenges}. With the use of privacy-preserving authentication on the rise, its emergence as a new CC category is expected \cite{beckers2012foundation}. For example, to ensure compliance with legislation (e.g., the European Union's General Data Protection Regulations), the feasibility of specifying the concept of privacy-by-design in the CC needs to be investigated, which can be achieved by incorporating privacy-preserving authentication into standards.

\textbf{Artificial intelligence systems:} Artificial Intelligence (AI) systems demonstrate the capability to produce tertiary consciousness such as self-recognition, cognitive feedback, components of self-concept, and so forth \cite{pagel2014dream}. However, in light of this capability as well as its misuse, the assurance of cyber security for AI systems is imperative. In January 2017, a group of artificial intelligence researchers developed 23 principles for AI called Asilomar AI principles \cite{AIprinciple}, which underlines that AI systems should be safe and secure throughout the whole operational lifetime, and verifiable so where applicable and feasible. Since then, many other questions remain as to what is a safe and secure AI system and how to achieve it, especially verifying the security features in the context of the rapidly expanding and developing areas in financial trading, health care, translation, transportation etc. The increasing dependence on AI for critical functions and services create more incentives for attackers and lead to more severe damages \cite{li2018cyber}. In the past years, there has been an accelerated growth of policy proposals and government interest in the security of AI system. For instance, the European Commission is proposing for a regulation laying down harmonized rules on AI \cite {AIproposal}. These security-focused policies for AI systems mainfest the importance of transparency, measurement and accountability for the AI system. 

\textbf{Internet of Things applications:} The IoT refers of the system of interrelated and internet-connected devices that are able to collect and transfer information over the wireless network without human to human or human to computer interaction \cite{fathy2018large}. With the emergence of IoT, numerous gadgets, services, and products containing innovative IoT technologies (such as smart TVs, voice controllers and mobile robots) provide consumers with convenience and improve the quality of their life \cite{kang2017obtain}. IoT ecosystems are complex with significant security challenges, including insufficient data protection, weak password protection, insecure interfaces, and other risks. However, these risks must be reduced and mitigated for the entire lifecycle of IoT devices. Recently, there are a few studies on the approaches to the CC certification for IoT devices  \cite{kang2017obtain} \cite{kim2017design} \cite{matheu2020survey} \cite{suciu2019lego} with a number of PPs developed in this space. Building trust in IoT devices with powerful IoT security solutions that keep IoT systems safe and ensures the availability, integrity, and confidentiality of the IoT solution is a potential future direction for the CC development. 

\begin{figure*}[ht]
    \centering
    \subfloat[\centering The number of Certified Products and archived Certified Products by TOE category (a Certified Product may have multiple categories associated with it)]{{\includegraphics[width=13cm]{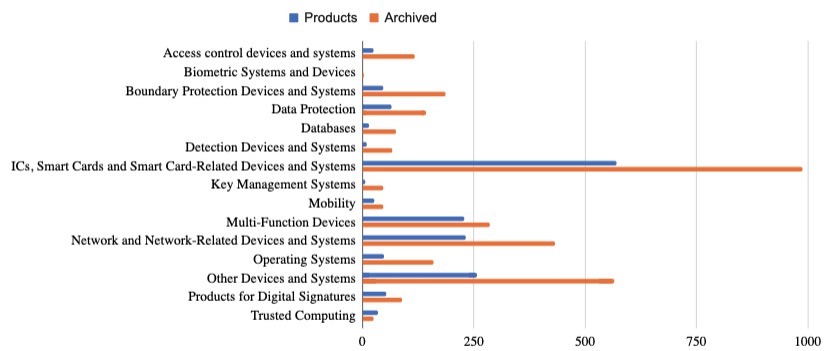} }}%
    \qquad
    \subfloat[\centering The number of Protection Profiles and archived Protection Profiles by TOE category (a Protection Profile may have multiple categories associated with it) ]{{\includegraphics[width=13cm]{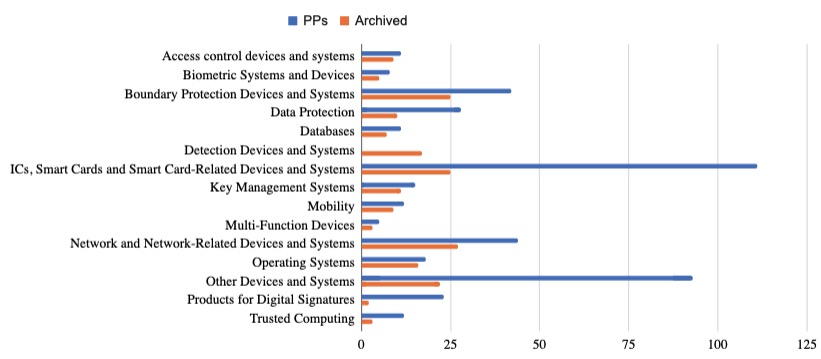} }}%
    \caption{The number of Certified Products and Protection Profiles under the Common Criteria certification scheme}%
    \label{fig:PP}%
\end{figure*}
\section{Common Criteria adoptions}
To pave the way for widespread adoption of the CC, we investigate possible adoption barriers to determine if organizations have concerns related to cyber security regulatory issues as well as deciding organizations' attitudes towards cyber security standards. In this section, we present the current worldwide adoption of the CC and identify the adoption barriers.  

\subsection{Trends in Common Criteria evaluations: some statistics}
To investigate the current adoption, development trend and users' trust of the CC internationally and nationally, we studied the cases, including Certified Products \cite{CC3} and Protection Profiles \cite{CC_pp}, listed on the CC portal \cite{CC_portal}. This subsection presents the statistics of the CC obtained after processing accessible information on the CC portal and the comparative analysis of Certified Products and Protection Profiles by category, scheme and assurance level.

 Figure \ref{fig:PP}(a) shows the number of Certified Products and archived Certified Products approved through the  CC evaluation and certification process. They are listed on the CC portal according to the categories of the TOEs. There are a total of 1619 Certified Products and 3226 archived Certified Products up to Dec 2021. The category of ICs, smart cards and smart card related devices and systems is the undisputed leader. Given the mutual recognition by all CCRA signatory countries and the comprehensive coverage of technologies and security functionalities, the number of signatory countries is progressively increasing \cite{ccabout}. Also, the number of Certified Products is rising annually. In the year of 2019 and 2020, the number of Certified Products was 275 and 367 respectively. Usually, the certificates will remain on Certified Products List with a five years validity\cite{CC3}. The validity period may be determined by a scheme or a technical community that developed the PP. The validity period may impact a vendor's return on investment from a certification. The trend of PPs is similar to Certified Products when looking at the number of PPs and archived PPs, whose number increases year by year. Based on the number of PPs and archived PPs by TOE category as shown in Figure \ref{fig:PP}(b), the category with the most Certified Products and PPs listed is the ICs, smart cards and smart card-related devices and systems.

\begin{figure}
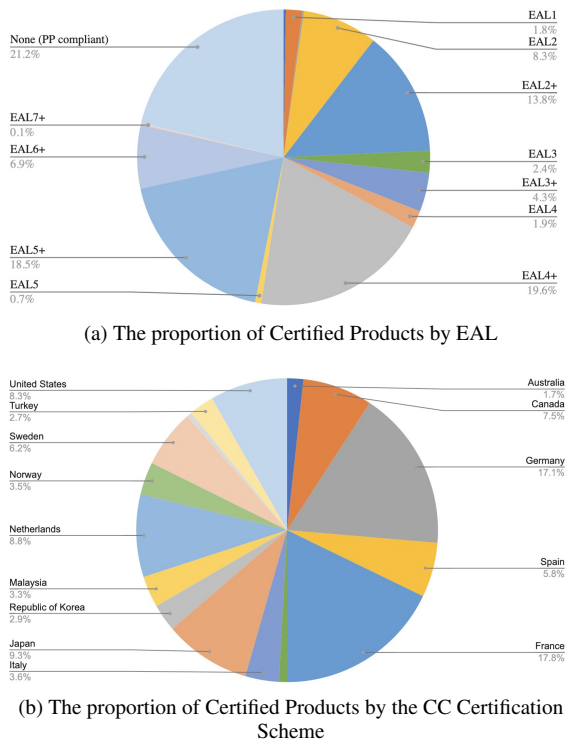
%
    \centering
    \subfloat[\centering The proportion of Certified Products by EAL]{{\includegraphics[width=7.5cm]{CP_byEAL.pdf} }}%
    \qquad
    \subfloat[\centering The proportion of Certified Products by the CC Certification Scheme ]{{\includegraphics[width=7.5cm]{CP_byScheme.pdf} }}%
    \caption{The proportion of Certified Products under Common Criteria evaluations }%
    \label{fig:proportionCP}%
\end{figure}

As mentioned above, the Evaluation Assurance Level is a number that ranges from EAL 1 to EAL 7, describing the depth and rigor of an evaluation.  EAL 1 is the least rigorous level, and EAL 7 the most exacting. Each EAL corresponds to a Security Assurance Requirements package that covers the complete development process with the given level of strictness. The Certificate Authorizing Participants implement CCRA compliant CC certification requirements to produce certificates under certificate authorizing schemes. Based on the 17 Certificate Authorizing Schemes \cite{CC_authority}, shares of Certified Products under different schemes, are shown in Figure \ref{fig:proportionCP}(b). As shown in Figure \ref{fig:proportionCP}(a), investigating by assurance levels, most products are rated under EAL4+. The French, German, and Netherlands schemes have produced the most certificates listed on the CC portal. When observing the number of Certified Products in 2021, the US leads the number on 33, Germany on 31, France on 20 and Canada on 8. In particular, the US keeps the Certified Products listed for two years before being archived, while the other countries keep them for five years. Different from the most applied schemes in Certified Products, it is worth noting that the United States scheme takes the maximum amount of adoption in PPs. In addition, in the same way of CC Certified Products, the EAL4+ takes up the largest proportion
(33.5\%) by investigating the PPs by certification assurance level. From the perspective of adoption of PPs, it is worth noting that the United States scheme certified against PPs exclusively. 

\subsection{Common Criteria adoption barriers}
\label{adoptionBarriers}
Despite the importance and value of the CC, there are barriers in it adoption. Below, we highlight the barriers to the adoption of the CC certification for ICT products.

\textbf{Cost and complexity of evaluation:}
    The CC evaluation cost is commonly regarded as a barrier to the CC certification adoption \cite{matheu2020survey} \cite{baldini2016security}, particularly for companies with a limited budget and for products with small market shares and low margin. In the competitive market of ICT products, products with a low-profit margin may not be able to justify and defray the cost of the CC certification. Based on the investigation on Certified Products \cite{CC3}, vendors of ICT products with high profit margin and capability for protecting sensitive government networks are relatively more inclined to adopt CC certifications. Extensive resources are required to complete the complex evaluation activities in the evaluation process by the laboratory. Based on the Australian scheme, the CC certification process generally consists of four phases, namely \textit{pre-evaluation}, \textit{conduct}, \textit{conclude}, and \textit{assurance continuity}. Generally, the pre-evaluation phase is essential to ensure success and avoid delays by conducting initial assessments and planning all building blocks of the CC evaluation process including development of the ST  and the schedule of evaluation. Other tasks include writing of the functional and high/low level design specifications. Secondly, the conduct phase is to verify any claimed security functionality under the requirements of the CC and other claimed cryptographic functionality under the specific security standard, such as FIPS 140-2 \cite{evans2002fips}. In the conclude phase, the evaluation and certification activities are finalised. Lastly, the assurance continuity phase establishes a way to minimize the number of evaluations and allows the determination to be made that the certification may be extended to the updated version of the TOE. The total cost of an evaluation varies depending on the security assurance level or PP conformance claims and the complexity of the TOE \cite{smith2007trends}. Four components, namely internal costs, external costs, lab fees, and certification fees make up the overall evaluation cost. The internal costs are incurred preparing deliverables and supporting the evaluators. The external costs consist of consulting fees. The lab fees are paid to the evaluation labs, and the certification fees are paid to the corresponding certification body if applicable. A recently estimated average cost for a CC certification lifecycle is US\$250,000 \cite{matheu2020survey} depending on the evaluation assurance level and re-use of past evaluation effort. The cost of an evaluation against a PP is generally lower. This is due to the reduced effort in developing the evaluation documentation. Because of the cost and complexity of evaluation, it is challenging to evaluate the products against the CC standard for companies with narrow profit margins and budgets.

\textbf{Time of evaluation:}
    With rapid changes in technology development, the time-consuming process of the CC evaluation and certification significantly slows down the commercialization of security products in markets. This is particularly undesirable for products with a short lifecycle. For example, the technology for producing a low-cost IoT device may have become obsolete when the CC certification process completes. The CC testing process needs to go through a sequence of stages, including ST evaluation, design evaluation, guidance evaluation, life-cycle evaluation, functional testing and penetration testing. Besides, the requirement on formal documentation and processes takes time \cite{kaluvuri2014quantitative} \cite{murdoch2012certification}. According to recent figures from the Australian Certification Authority, the average evaluation completion time is around six months. According to the Cyber Security Agency of Singapore, a typical evaluation takes about 3 to 6 months \cite{cybersecurityAgencyofSingapore}. According to \cite{tierney2017common}, a CC certification of an EAL 4 evaluation would take 6 to 9 months depending on the technology type. The completion of the CC certification can take about 9 to 12 months \cite{whatiscc},  according to the Lightship Sec Lab. The Canadian Centre for CyberSecurity would allow the total of about 6 months from the Product in Evaluation (PiE) date \cite{canadianCC}. Generally, the average time needed for the CC certification is 6 months to 1 year \cite{matheu2020survey}. For products that require frequent replacements or updates or have a short time-to-market, the time for evaluation indeed hinders the adoption of the CC evaluation.

   \textbf{Lack of government supports and incentives:}
    Government supports and incentives promote the adoption of the CC evaluation and CC Certified Products to a certain degree. The consumers of CC Certified Products can be categorized based on market sectors: the public and private sectors. In terms of the public sector, government support can increase the adoption of the CC certification by requiring the procurement of CC Certified Products. Furthermore, setting policy requirements for the procurement process used by governmental departments and agencies stimulates CC adoptions. For example, the US government issued a policy requiring the CC certification for products intended for certain applications, which encourages vendors to participate in CC evaluations \cite{smith2007trends}. With respect to the private sector, government incentives can promote the adoption of CC Certified Products as well. For instance, the IT Investment Promotion Tax Incentive in Japan allows businesses to claim tax deductions for the use of CC Certified Products, which consequently increases the acquisition of the Certified Products \cite{yajimaconsideration}. The government's support and incentives would play an essential role in boosting the adoption of CC certification as the vendors can address legal risks and gain economic benefit from performing the CC evaluation. 
  \begin{table*}[]
\centering
\caption{Comparison of state-of-the-art cyber security standards regarding scope (I indicates international standard, N indicates national standard, and D indicates industry-specific stand), mutual recognition agreement (Y indicates that more than two countries agree on the standard), target candidates, publicly available (Y means that the security standard is publicly available, N indicates standard is not publicly available), cost and time for evaluation (NA means the cost and time spent on the standard are not available)}
\label{standard_comparison}
\begin{tabular}{@{}lllllll@{}}
\toprule
Standard                                                                            & Scope & \begin{tabular}[c]{@{}l@{}}Mutual\\ Recognition\\ Agreement\end{tabular} & \begin{tabular}[c]{@{}l@{}}Target \\ Candidate\end{tabular}            & \begin{tabular}[c]{@{}l@{}}Publicly\\ Available\end{tabular} & Cost                                                                                                                            & Time                                                          \\ \midrule \midrule
\rev{CC}                                                                                & I     & Y                                                                        & Product                                                                & Y                                                            & US\$250,000                                                                                                                      & \begin{tabular}[c]{@{}l@{}}6 months \\ to 1 year\end{tabular} \\
\hline
\rev{ISO 27K}                                                                             & I     & Y                                                                        & Organization                                                           & Y                                                            & \begin{tabular}[c]{@{}l@{}}\pounds2,850 \\ to \pounds14,250\end{tabular}                          & \begin{tabular}[c]{@{}l@{}}2 to \\ 15 days\end{tabular}       \\
\hline

\rev{IEC 62443}                                                                           & I     & Y                                                                        & \begin{tabular}[c]{@{}l@{}}Industrial \\ Automation\\ and Control\\ Systems\end{tabular}                                                                & Y                                                            & \rev{\begin{tabular}[c]{@{}l@{}}US\$500 to\\ US\$20,000\end{tabular} }                                                                                                                              & NA                                                            \\
\hline
\begin{tabular}[c]{@{}l@{}}\rev{ISO/SAE} \\ \rev{21434}\end{tabular}                            & I     & Y                                                                        & \begin{tabular}[c]{@{}l@{}}Automotive\\ Industry\end{tabular}          & Y                                                            & NA                                                                                                                              & NA                                                            \\
\hline
\begin{tabular}[c]{@{}l@{}}\rev{ISO/IEC} \\ \rev{20924} \\ \rev{\&} \rev{ETSI EN} \\ \rev{303645}\end{tabular} & I     & Y                                                                        & IoT                                                                    & Y                                                            & NA                                                                                                                              & NA                                                            \\
\hline
\begin{tabular}[c]{@{}l@{}}\rev{Cyber} \\ \rev{Essentials}\end{tabular}                         & N     & UK                                                                       & Organization                                                           & Y                                                            & \pounds 300                                                                                                      & \begin{tabular}[c]{@{}l@{}}1 to\\ 14 days\end{tabular}        \\
\hline
\rev{Essential 8}                                                                         & N     & Australia                                                                & Organization                                                           & Y                                                            & NA                                                                                                                              & NA                                                            \\

\hline
\begin{tabular}[c]{@{}l@{}}\rev{UK NCSC} \\ \rev{\&} \rev{CPA}\end{tabular}                           & N     & UK                                                                       & Smart Meters                                                           & Y                                                            & \begin{tabular}[c]{@{}l@{}}US\$1,300 \\ per day\end{tabular}                                                                     & \begin{tabular}[c]{@{}l@{}}6 to \\ 18 months\end{tabular}     \\
\hline
\rev{CSPN}                                                                                & N     & France                                                                   & Product                                                                & Y                                                            & \begin{tabular}[c]{@{}l@{}}€ 25,000\\ to € 35,000\end{tabular}                        & \begin{tabular}[c]{@{}l@{}}35 days\\ to 2 months\end{tabular} \\
\hline
\rev{IT-Grundschutz}                                                                      & N     & German                                                                   & Organization                                                           & Y                                                            & NA                                                                                                                              & NA                                                            \\
\hline
\rev{NIST}                                                                                & N     & US                                                                       & Organization                                                           & Y                                                            & \rev{\begin{tabular}[c]{@{}l@{}}US\$5,000 to\\ US\$15,000\\ on compliance \\check,  \$35,000 to\\ \$115,000 on \\remediation once \\issues found  \end{tabular}     }                                                                                                                    & \rev{\begin{tabular}[c]{@{}l@{}}A few weeks\\ to several years          \end{tabular} }                                                \\
\hline
\rev{NERC CCS}                                                                            & N     & \begin{tabular}[c]{@{}l@{}}North\\ American\end{tabular}                 & \begin{tabular}[c]{@{}l@{}}Electrical \\ Power\\ Industry\end{tabular} & Y                                                            & NA                                                                                                                              & NA                                                            \\
\hline
\rev{FIPS 140}                                                                            & N     & US                                                                       & Cryptography                                                           & Y                                                            & \begin{tabular}[c]{@{}l@{}}US\$50,000 per\\ module\end{tabular}                                                                   & \begin{tabular}[c]{@{}l@{}}Up to \\ 1 year\end{tabular}       \\
\hline
\rev{PCI DSS}                                                                             & D     & Y                                                                        & \begin{tabular}[c]{@{}l@{}}Payment \\ Card\end{tabular}                & Y                                                            & \begin{tabular}[c]{@{}l@{}}US\$15,000\\ to US\$40,000\end{tabular}                                                                  & NA                                                            \\
\hline
\rev{UL 2900}                                                                             & D     & Y                                                                        & \begin{tabular}[c]{@{}l@{}}Medical \\ Device\end{tabular}              & N                                                            & \begin{tabular}[c]{@{}l@{}}US\$225 to \\US\$750 on checking \\ standards, \\ US\$40,000 to\\ US\$150,000 on\\ certification\end{tabular}& NA                                                            \\ \bottomrule

\end{tabular}
\end{table*}
  
  \textbf{Availability of approved Protection Profiles:}
    The CC certification process of ICT products may be hindered due to the lack of approved PPs. For example, the Australasian and Singaporean CC schemes encourage products to be certified against an approved PP \cite{cybersecurityAgencyofSingapore}\cite{AISEP}. Products to be certified without PP conformance (such as using ST only) may only be accepted on a case-by-case basis or when no suitable PP exists. Hence, the unavailability of approved PPs may discourage vendors from obtaining the CC certification. For mutual recognition under the CCRA, a CC certificate claiming compliance to EAL 3 or higher but not claiming compliance to a collaborative Protection Profile is generally treated as an equivalence to EAL 2 \cite{cybersecurityAgencyofSingapore}. 

  \textbf{Implementation of security-by-design in products:}
    Implementing \textit{security-by-design} in product engineering processes means that the product must be designed from the ground up to be secure. This may shorten the evaluation and certification process significantly \cite{cybersecurityAgencyofSingapore} because the incremental certification of products for additional product features can be more accessible with the integration of certification evaluation activities in the security system engineering process \cite{andrea2020towards}\cite{beznosov2004towards}. However, it is challenging to integrate \textit{security-by-design} in security product engineering processes \cite{peisert2014designed}. Therefore, the lack of \textit{security-by-design} in products may leave the products with many vulnerabilities \cite{patton2014uninvited}, which makes it difficult for obtaining a security certification.

   \textbf{Complexity of the CC standard:}
    The structure and the readability of the CC standard is somewhat complex and not easily understandable. This has led to the need for many product vendors to engage evaluation supporting consultants at the pre-evaluation stage in order to prepare specific evaluation material which increases the overall evaluation time and cost. 
    Major review work is underway by international experts through the International Organization of Standardization (ISO). This should see an improvement of the CC for wider adoption. Further review work is likely needed to ensure usability and readability and to make the standard simpler to understand and use.

\subsection{State-of-the-art of security standards}
\label{securityStandards}

Besides the CC, there are also other cyber security standards that are designed to protect the cyber operational environments of users and the organizations involved \cite{pillitteri2014guidelines}. Starting from the objective that mitigates and reduces security risks, a series of cyber security standards are promulgated, including policies, guidelines, best practices, and so forth, to contribute to establishing a trusted cyber security environment. This subsection surveys the currently existing and state-of-the-art standards, including international standards, national standards, and industry-specific standards, which are widely considered to address the cyber security posed by threats. To further demonstrate the impacts of adopting the CC, Table \ref{standard_comparison} summarizes the state-of-the-art cyber security standards and compare the CC with these cyber security standards from the perspective of the scope, mutual recognition agreement, target candidates, publicly available, and cost and time for evaluation.

\subsubsection{International Standards}
 We first outline the international standards, focusing on the objective, context of use, and cost and time for obtaining the certificate. 
\rev{
\textbf{ISO 27K:} ISO/IEC 27000-series, also known as Information Security Management System (ISMS) Family Standards or ISO 27K, consists of information security standards published by International Organization for Standardization (ISO) and the International Electrotechnical Commission (IEC). The representative ISO/IEC 27001 establishes requirements on how to manage information security \cite{humphreys2016implementing}. The objective of ISO/IEC 27001 is to provide guidelines for organizations on how to manage the information and data. Therefore, the standard is not applicable to the IT industry only, also to other candidates, including organizations and government bodies that aim to protect their information. Through the requirements for establishing, implementing, maintaining and constantly improving the ISMS, the standard helps organizations make data assets more secure \cite{humphreys2016implementing}. The recognized national accreditation body properly accredits the certification if the organizations meet the requirements and pass the audits, which usually costs from \pounds2,850 to \pounds14,250 and takes 2 to 15 days \cite{ISO27001} based on the size of the organization. Furthermore, ISO/IEC 27002 provides best practices on information security management for those who are responsible for implementing and maintaining the ISMS. Besides, ISO 27K includes the standards from ISO/IEC 27000 to ISO/IEC 27050, ISO/IEC 27701 and ISO/IEC 27799 \cite{higgins2009information}. The ISO 27K standards are routinely reviewed and updated on a roughly five-year cycle. The standards related to digital forensics and cyber security are in preparation. Compared with the CC, ISO 27K is related to the information security certification for companies, while the CC certifies products.

\textbf{IEC 62443:} Some standards tend to specialize in a specific domain in cyber security. IEC 62443 targets Industrial Automation and Control Systems (IACS) by defining common standards in processes, techniques and security requirements \cite{leander2019applicability}. There are four categories in IEC 62443 cyber security standard series, respectively General, Policies and Procedures, System and Component, covering foundational information, asset owner, system design guidance and requirements, and specific product development and technical requirements for IACS \cite{hauet2012isa99}. IEC 62443 defines Common Component Security Constraints (CCSCs) in addition to technical requirements, which must be met by components to be compliant with IEC 62443 Part 4.2.  
   
 \textbf{ISO/SAE 21434}: New vehicle usage trends, including car-sharing platforms and mobility-as-a-service, are growing. However, the number of attack vectors in the connected cars and automotive industry is significant \cite{ISO21434}. Under the circumstance, a new cyber security standard for the development lifecycle of road vehicles is under development by ISO and the Society of Automotive Engineers (SAE). ISO/SAE 21434 sets the guidelines for securing the high-level processes and cyber security standards in the connected cars and automotive industry \cite{schmittner2018status}. Different phases, namely requirements engineering, design, specification, implementation, test and operations are taken into consideration on security aspects. 
   
\textbf{ISO/IEC 20924 and ETSI EN 303645:} During the past decade, IoT that encompasses various protocols and technologies to interconnect physical devices to Internet infrastructure is one of the most relevant scenarios in cyber security \cite{al2015internet}. ISO/IEC 20924 \cite{ISO20924} provides the definition along with the terms and definitions forming the terminology foundation for IoT, which fills the gap between traditional security standards to the extension on the adoption of IoT systems in the early stages \cite{granjal2015security} \cite{makhdoom2018anatomy}. Furthermore, the ETSI EN 303645 standard provides the baseline requirements for security in IoT devices \cite{303645}, which was released in 2020. The standard compasses the technical controls and organizational policies for the developers and vendors of IoT devices \cite{303645}. For IoT devices involving multiple Personally Identifiable Information (PII), the standard facilitates their compliance to the General Data Protection Regulation (GDPR) \cite{regulation2018general}.

\subsubsection{National Standards}
Besides international security standards, national governments enact standards to set expectations and requirements for ICT products and organizations regarding cyber security. The national standards are described below with a focus on the application context, harmonization with international standards, and evaluation workflow.

\textbf{Cyber Essentials:} Cyber Essentials is a national security standard developed by the UK government that specifies the assurance framework and a set of security controls to protect information from threats, concerning technical rules designed to protect devices, internet connection, data and services \cite{ calder2014cyber}. This standard is a government-backed certification scheme that helps to tighten overall cyber security within the organization, and it is also mandatory for businesses looking for specific government contracts \cite{essentials2015cyber}. The certification usually takes a day to a fortnight to complete the assessment and costs around \pounds 300. Cyber Essentials certification is valid for 12 months upon successful application.

\textbf{ASD Essential 8:} The Essential Eight is a series of baseline mitigation strategies to combat cyber security incidents produced by the Australian Signals Directorate (ASD) \cite{essential8}. These strategies aim to aid organizations in protecting against adversaries to compromise systems. The Essential Eight Maturity Model provides advice on how to implement the strategies as well as assists the organizations in self-testing the maturity of implementation. As one of the most effective mitigation strategies in ensuring the security of systems, application control is designed to protect against malware executing on the systems. The other strategies include assessing security vulnerabilities and applying patches, Microsoft Office Macro Security, restricting administrative privileges, implementing multi-factor authentication, and the other strategies to mitigate cyber security incidents.  

\textbf{UK NCSC and CPA:} National Cyber Security Centre \cite{NCSC} is the UK government organization that gives support and advice to the public and private sector to help them avoid security threats. The Commercial Product Assurance (CPA) \cite{CPA} was set up to help organizations demonstrate the security functions of the products met defined NCSC standards. The Assured Service Providers is used to conduct testing and assessment by NCSC. The products are tested against the published CPA Security Characteristics \cite{CPA}. The certification usually needs 6 to 18 months, and the cost of the certificate is US\$1,300 per day of work \cite{JRC}. However, since March 2019, the NCSC no longer accepts new product evaluation under the CPA scheme unless they are Smart Meters or smart metering products. 

\textbf{CSPN:} The Certification de Sécurité de Premier Niveau (CSPN) is a cyber security certification methodology proposed by the National Cybersecurity Agency of France (ANSSI) in 2008 \cite{CSPN}. The main objective of CSPN is to verify the product's compliance with its security specifications. Compared with the CC, CSPN is a lightweight certification standard \cite{matheu2020survey} that assesses the product in a shorter period, typically taking from 35 days to 2 months. Generally, the cost for evaluation and certification is around from €25,000 to €35,000 \cite{JRC}. However, CSPN is only recoginzed in France as a national standard. 

\textbf{BSI IT baseline protection:} IT baseline protection (also known as IT-Grundschutz) from the German Federal Office for Information Security (BSI) is a methodology to identify and implement cyber security measures in an organization. The goal of BSI IT baseline protection is to achieve the adequate and appropriate level of security for IT systems \cite{tsoumas2006towards}. To achieve the goal, BSI recommends ``well-proven technical, organizational, personnel, and infrastructural safeguards" \cite{BSI}. The organizations show the systematic approach to secure their IT systems by obtaining the ISO/IEC 27001 certificate based on IT-Grundschutz.

\textbf{NIST:} The National Institute of Standards and Technology (NIST) of the US proposed cyber security framework on the basis of standards and guidelines to help organizations manage security risks \cite{NIST}. Based on Executive Order 13636, the framework was defined to improve the cyber security of critical infrastructure in 2014 \cite{NIST1} and updated to address the emerging scenarios in cyber security in 2018 \cite{NIST}. Furthermore, a series of special publications describe the security principles and provide advice on cyber security management, which is aligned to the framework.

\textbf{NERC CCS:} Some national standards target a specific area in cyber security. North American Electric Reliability Corporation (NERC) Cyber Security Standards (CCS) was created in 2003 to be used to secure the electrical power industry \cite{NERC}. As the enhancement of the requirements, the newest and most widely recognized NERC security standard is NERC 1300, which provides bulk electric system standard to network administration and supports the best-practice industry processes.

\textbf{FIPS 140:} In the area of cryptography and database security, Federal Information Processing Standards (FIPS) are the US government security standards that specify the requirements for the cryptography modules \cite{standard1995fips}. FIPS 140-2 and FIPS 140-3 are current and active standards. In support of the cryptography block in CC functional security requirements, FIPS provides the specifications for cryptographic modules, and a set of standards that specify the cryptographic algorithm in use \cite{brown1994security}. Four security levels from Level 1 to Level 4 are defined in FIPS 140-2. FIPS 140 validations can take up to one year and cost over \$50,000 per module. The individual ratings and overall rating are listed on the vendor's validation certificate in the general flow of FIPS testing and validation \cite{no1fips}.

\subsubsection{Industry-specific Standards}
After recapping the international and national standards, we further replenish the state-of-the-art security standards by including the industry-specific standards.

\textbf{PCI DSS:} The usage of payment cards, such as debit card, credit card, and prepaid card, is continuously increasing \cite{liu2010survey}. Consequently, the number of security incidents, like data breach, related to payment cards cause damages on businesses, customers' benefits and retailers' reputation \cite{morse2008pci}. The Payment Card Industry Data Security Standard (PCI DSS) is an industry-specific security standards administered by the PCI Security Standard Council, which sets the information security standards for organizations that handle payment cards from the major card schemes. The objective of the standard is to tighten payment card information and reduce cyber incidents on payment card, such as credit card fraud. An audit that accesses the organization's compliance with the PCI DSS costs around \$15,000 to \$40,000, depending on the business types, size, security environment, and the specific processing methods of the usage of payment card \cite{ataya2010pci}.

\textbf{UL 2900:}
UL 2900 is a set of standards published by Underwriters Laboratories (UL) that is a global security certification organization. And it includes the general cyber security requirements (UL 2900-1), specific requirements for medical products (UL 2900-2-1), industrial systems (UL 2900-2-2), and security and life safety signaling systems (UL 2900-2-3) \cite{yuan2018standards}. 
The standard requires effective security countermeasures implemented to protect data as well as other data assets, such as command and control data. Besides, security vulnerabilities in the software should be eliminated, and the security of software should be verified through penetration testing \cite{ULstandards}. The UL 2900 standards are not publicly available, leading to the harmonization and standardization aspects that should be addressed further \cite{matheu2020survey}. Typically, check on the criteria will spend between US\$225 to US\$750 based on the number and delivery format of the standards. And a fee ranging from US\$40,000 to US\$150,000 will be needed to certify the products \cite{UL}. 
}

\section{Defining security requirements with the common criteria}

In this section, combined with the lessons learned from the recent project, \textit{Development of the Australia Cyber Criteria Assessment (DACCA)}, we present the lessons learned from and reflect on the challenges of defining security requirements with the CC through the development of a Protection Profile. In addition, we propose potential future directions to improve and promote the adoption of the CC. 

\subsection{Best practices}

\begin{figure*}
  \centering
  \includegraphics[width=0.9\linewidth]{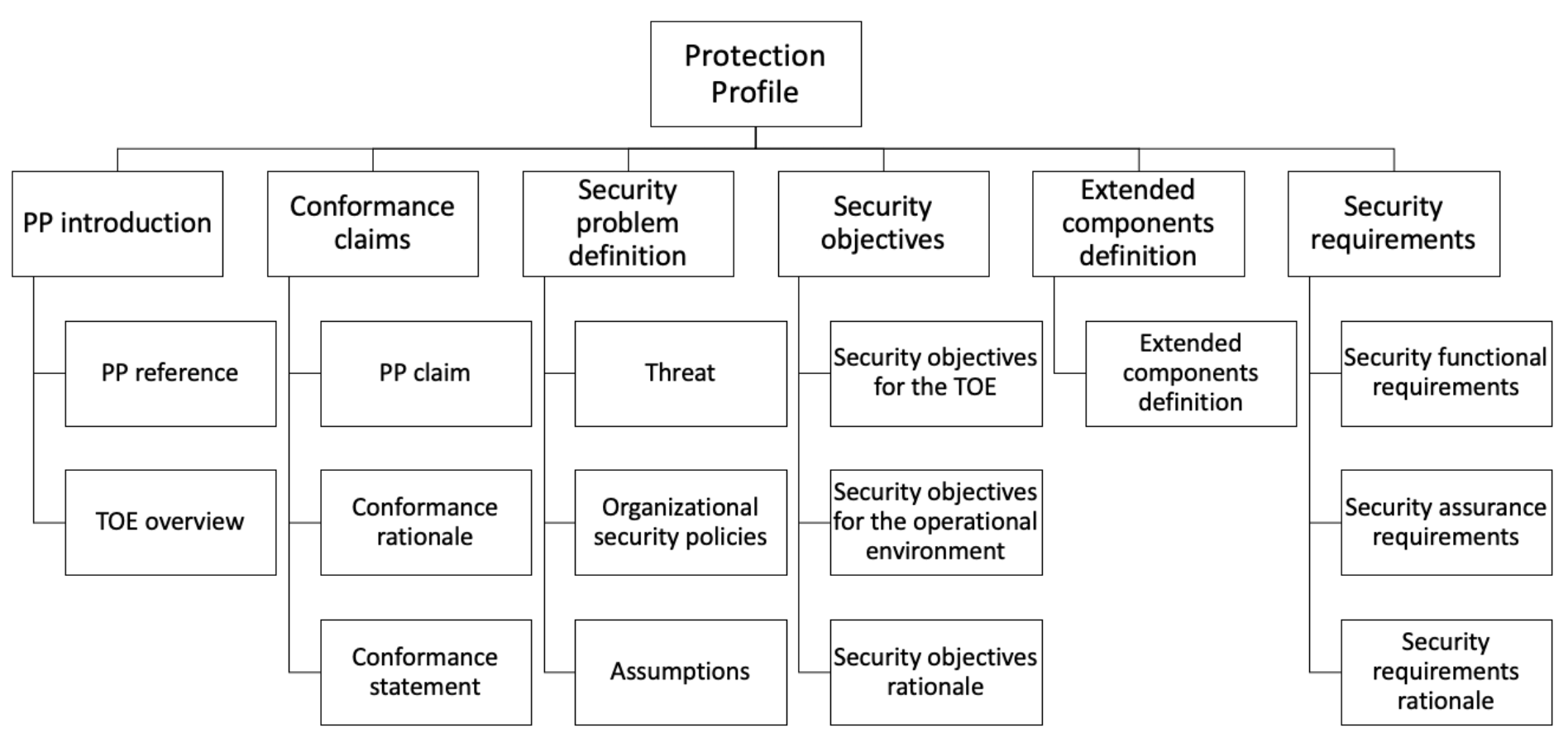}
  \caption{Main structural blocks of a Protection Profile}
  \label{ppspecifications}
\end{figure*}

\rev{In Figure \ref{ppspecifications}, the structure of a generic PP is outlined, which consists of the main blocks, including PP introduction, conformance claims, security problem definition, security objectives, extended components definition, and security requirements. In terms of the development process of a PP, it is an iterative and incremental procedure where the specifications are broken down into multiple blocks. The development process is gradually and iteratively built up with the core blocks, and the supplementary features are added further.} \rev{In practice, we developed a PP to define security requirements with the CC for the target TOE - encryption key managers. The following are the best practices we summarized on how to define security requirements with the CC in a Protection Profile through the literature review and the experience gained from the \textit{DACCA} project.}

\textbf{Iterative and incremental approach:} \rev{We have taken an incremental approach in the project by iterating the following steps to mature the PP. This approach proved to be of particular effectiveness for an inter-sectoral project like \textit{DACCA} with partners from the academia, industry, and certification authority. The first step, \textit{Q\&A}, begins with brainstorming with initial questions and answers. The initial questions identified to help explain the TOE and the objectives include what the common characteristics and functions for the encryption key managers are, what the required non-TOE hardware/software/firmware for the TOE is, what the operational environment for the encryption key management products is, and so forth. The second step, \textit{Draft}, aims to draft contents for each chapter and section following the PP specifications listed in Figure \ref{ppspecifications}. The third step, \textit{Refinement}, refines the contents based on literature review, cross-reference to related PPs, and feedback from industrial partners. The fourth step, \textit{Polishing}, is to reduce impractical and repetitive items surrounding the risks, which makes sure the security countermeasures are correct and sufficient to respond to the considered risks. The fifth step \textit{Evaluation} reviews and evaluates the deliverability of the PP to improve the consistency with the CC standard. The incremental process of PP developments iterates to gradually improve the quality of PP documents and the assurance of security.}
    
\textbf{PP-Module development:} \rev{The new evolution of the CC supports the comparability among the results of independent cyber security evaluations through collaborative Protection Profile (cPP) \cite{matheu2020survey}. A PP-module builds on a cPP, and conforming TOEs are obligated to implement the functionality required in the cPP along with the additional functionality defined in the PP-module. Hence, building upon the cPP rather than developing a standalone PP will not reinvent the wheel on certain functionality and ensure the specified functionality is sufficient to enhance cyber security for the technology and product. For example, in the \textit{DACCA} project, for the target TOE - key encryption management components, we utilize the collaborative Protection Profile for Network Devices (NDcPP) as the Base-PP and develop PP-Module intended for use with the NDcPP. This Base-PP is valid because a device that implements centralised enterprise Encryption Key Management is a specific type of network device. There is nothing about implementing Encryption Key Management that would prevent any of the security capabilities defined by the Base-PP from being satisfied. In the developed PP-Module, only the security functionalities of Encryption Key Management components in terms of the CC and assurance requirements for such products that are not included are needed to be added in the PP-Module, which avoids repetitive and redundant information, and also simplifies the development process.}
    
\textbf{Threats analysis:} \rev{A CC certification requires comprehensible product documentation, including a detailed threat analysis \cite{beckers2013problem}. The following analytical approaches have been proved to be effective for such a purpose in our PP development. The anecdotal way is generally adopted by brainstorming a list of known threats and then culling, categorizing by assets, and assigning to operating environments. An alternative and more analytical way is to order, delete or retain the threats according to the priority, severity, and mitigation cost of the threats. Alternatively, work in \cite{brian} defined the data as assets and then generalized data assets (e.g., documents, configuration data) and asset state (e.g., in transit, at rest) to apply different threats to different states. Different threats may materialize in various operational environments or arise from vulnerabilities. Hence, defining the security threats based on the operational environments and supplementary threats derived from various vulnerabilities are another two effective approaches. The hybrid method that combines the above five approaches makes sure the identified threats in the PP are comprehensive, which paves the way for specifying the security requirements desired for the product under the CC standard.}

\textbf{Security functional requirements specification:} \rev{Specifying high-quality security requirements is an essential but complex task \cite{mellado2007common}. The starting point can be  tailoring through existing SFRs listing in the CC Part 2 \cite{CCpart2} through \textit{iteration, assignment, selection and refinement}. Generally, \textit{iteration} enables a component to be used more than once with varying operations. \textit{Assignment} provides the specification of parameters. \textit{Selection} allows the specification of one or more items from a list. \textit{Refinement} permits the addition of details. However, there are some security objectives for the TOE that cannot be transformed to SFRs listed in the CC Part 2, for example, organizational policies or other third party requirements. The extended security functional requirements are set and included with definition and detailed requirements on the condition that the security objectives are hard to translate. Furthermore, in the PP development process, the SFRs are required to be specifically and precisely defined based on the corresponding functionalities of target TOE instead of generically described. For example, when defining the SFRs of key encryption management components in the \textit{DACCA} project, cryptography's concrete aspects, including the cryptographic features of encryption and decryption, hashing, keyed hash algorithm, and digital signatures, are included in the SFRs rather than the general description.}

\subsection{New challenges ahead}
 Informed by the best practices we have presented, and in the light of the security implications attributed to emerging technologies, we highlight the following challenges that will be addressed for the development and broad adoption of the CC.

\subsubsection{Transferability of Protection Profiles} 
With the target TOE, there are many different configurations, methods and parameters for key management products \cite{ghosal2019key}\cite{daniel2014emerging}. For example, a pure software key management product is a standalone TOE that generates, stores and manages key materials. Moreover, some key management products use specialised hardware for key material generation through True Random Number Generators (TRNGs), or for providing extra security properties via an internal Hardware Security Module (HSM). Besides, some key managers require the operating system for the supporting hardware platform or have the embedded operating system in the product. To fulfill encryption-related operations, such as using key material to sign by clients, the key management products require client-side software to be securely linked. The description of the TOE is the foundation of a PP and determines the number of PPs that will be developed \cite{vorobiev2010ontology}. To achieve harmonization, the TOE description facilitates the further comparison of the same types of products when conducting evaluation by comparing technical documents \cite{cugini1995common}. \textit{Can one Protection Profile satisfy a class of products that is composed of many combinations of options} is a challenge to be addressed in future PP development.

\subsubsection{Compactness and Sufficiency of Requirements} The CC has already defined 11 categories of security functional requirements concerning desirable security functionalities to provide a standard way of expressing the requirements for a TOE. However, the up-front analytical work is essential when specifying the complete, atomic, and testable requirements through the lifecycle for gathering, analyzing, and synthesizing security requirements \cite{brian}. In the lifecycle of PP development, the SFRs will be refined by filtering the irrelevant requirements and noisy features. Therefore, another main challenge is \textit{the formulation of a compact but sufficient set of security functional requirements}. Additionally, the SFRs need to be polished through matching and augmenting by packages for each function and option of key management products, as mentioned in the first challenge.
    
\subsubsection{Future-proofing of Requirements} Due to the complexity and diversity of operational environments, the CC approach might lack linguistic expressiveness for the full range of security requirements. The CC is technology agnostic in terms of security functional and assurance requirements and this leads to the formulation of product-specific requirements. Working in collaboration with industry partners to identify relevant product and system specific requirements for inclusion in the PPs is important. Simultaneously, PP developers need to ensure that the requirements do not contradict the CC requirements. Additionally, adding threat awareness into the PPs by incorporating a better understanding of threats from vulnerability is desirable in the PP development \cite{ardi2009introducing}\cite{li2017fesr}. Furthermore, the security implication of emerging technologies, such as quantum computing and zero-knowledge proofs, need to be considered in the future PP development \cite{farkas2002perspective}. In summary, extended requirements are expected to be incorporated based on analysis of features of products, security threats, and emerging technology. Extrapolation from existing guidance and Protection Profiles for complicated operational environments is the direction for extensions of CC requirements.

\subsection{Future directions} 
Below, we summarize the key findings from our study of CC adoptions and the PP development with the CC. Moreover, the recommendations and future directions for the CC to establish the trusted security ecosystem are proposed.

\textbf{Latest renovations of the Common Criteria:}
There is the new generation of PP type where an EAL is not specified within the PP itself. That means assurance is gained through customized assurance activities developed as part of the PP for the given technology and is based on SARs of different assurance levels. Another renovation is the transformation of evaluations with exact compliance to technology-specific PP to provide achievable, repeatable, testable evaluation results.  For example, the National Information Assurance Partnership (NIAP) \cite{NIAP_portal} which oversees a national program to evaluate Commercial Off-The-Shelf (COTS) ICT products' conformance to the CC, no longer accepts EAL-based evaluations. Products being evaluated against a NIAP-approved PP must be in exact compliance with that PP. NIAP has worked closely with government agencies, including the National Institute of Standards and Technology (NIST), to ensure all references to Evaluation Assurance Levels and Robustness were removed from applicable documentation. Occasionally, EAL or Robustness is mentioned, usually in regards to product acquisition.

\textbf{Accommodation of emerging technologies:} The CC covers a wide range of ICT security-related technologies and the evaluations in terms of security functionalities and security assurance. In Section \ref{current_categories}, the categories of existing CC applications were reviewed. Traditional ICT technology and products, such as ICs, database and network devices, are sufficiently covered and evaluated under the CC standard in the past decades \cite{herrmann2002using}. In light of emerging technologies as listed in Section \ref{potential_categories}, the evaluation of the newly emerging technologies, including blockchain, quantum computing, AI, and IoT, as well as the compliance requirements with privacy legislations for high assurance products, such as privacy-preserving authentication, need to be covered in the CC standard.

\textbf{Elimination of Common Criteria adoption barriers:}
    Increased adoption of the CC evaluation contributes to improving ICT ecosystem security for end-users. To lay the foundation for more extensive adoption of the CC evaluation, we analyzed the CC adoption barriers in Section \ref{adoptionBarriers}. An effective way to bolster the CC adoption is to eliminate the identified barriers. The possible solutions include cost, time and complexity reduction through the normalization of evaluation process, support and incentives from government agencies, increase of the availability and coverage of cPPs, and the integration of evaluation activities with product design and engineering procedures.

\textbf{Standardization of evaluation activities:}
    Traditional PPs are implementation-independent documents \cite{lee2010protection}, which define the security requirements for the ICT technology that the consumers require. Competent and independently licensed laboratories evaluate the products to decide whether the claimed security properties have been achieved\cite{CCevlauationlab}. For cPPs, unified processes and formalized steps for evaluation activities reduce the complexity of the evaluation. The standardization of evaluation and testing procedures improves the transparency of the certification process. Through standardization, the evaluation activities are better defined in PPs and corresponding supporting documents to provide guidance for evaluators.

\textbf{Comparability and harmonization of evaluation:} Based on the fact that various cyber security standards internationally, nationally, and at the industry-specific level are adopted worldwide, it is hard to compare the level of security properties across diverse security standards. Even for the CC standard, it is difficult to reach the objective of comparison due to complex technical documents \cite{matheu2020survey}. The standardization process of evaluation activities mentioned above can improve the comparability and harmonization of evaluation. In addition, rigorous security metrics that indicate the level of threats, risks and security provided by the products can be developed to address the issue of comparability.
     
\textbf{Improvement of user confidence:} The adoption of security-sensitive ICT products and services heavily relies on the users' trust in the security functionality of the ICT products. Cyber security standards and certification is considered as a driving force for increasing the users' trust. The collaboration among vendors, technical specialists, customers and governments to raise the bar of security in ICT products is a never-ending endeavor. The sharing of information on the core blocks of the CC evaluation and certification, such as threats, vulnerabilities and evaluation activities through education and information provided on the CC portal \cite{CC_portal} or other platforms will also contribute to this cause. Similarly, improvement to the usability and readability of CC standards would also help with the users of the standard. 

\rev{\textbf{Effective tools for security specifications:} CC security specifications are written in natural language, making rigorous evaluation challenging. As part of the new evolution of the PP, all evaluation activities are thoroughly defined in the corresponding supporting document. Furthermore, some tools have been developed to fulfill security requirements. For example, Morimoto et al. \cite{morimoto2007formal} proposed a process that makes the security specifications with the CC formalized in a mathematics manner. In addition, Teri et al. \cite{teri2000using} introduced a model called B method that can formally model security specifications of the Java Card. There is a high demand for effective tools that assist in making the evaluation process efficient, reliable, and rigorous.}

\section{Conclusion}
The last decade has witnessed a security paradigm shift from subjective risk management to more objective and measurable trust validation. One of the main driving forces of this shift is security evaluation and certification. The promotion of security standards, particularly the CC, facilitates mutual recognition of secure ICT products and adds value to the products and services to give the industrial partners the competitive edge to operate in the global market. Certification provides consumers with a level of assurance in the security of ICT products and enables them to make better-informed decisions when it comes to procurement. This paper provides the readers who are interested in trusted cyber security development from academia, government agencies, and industry with references and guidelines for the specification, development, evaluation, certification, procurement, and operation of ICT products with security functionality. This paper provides a rigorous review of the CC by summarizing the methodology of the CC, analyzing CC applications, investigating CC adoptions, and comparing the CC with the state-of-the-art cyber security standards. Through our experience from the PP development of an inter-sectoral project, we presented lessons learned from defining security requirements through the Protection Profile. We identified the challenges of defining security requirements through the CC and offered suggestions on the direction of defining security requirements for trusted cyber security.

\section*{Acknowledgment}{We acknowledge Mr Bosheng Yan's contribution to data collection in relation to Protection Profile development through his internship associated with the \textit{Development of Australian Cyber Criteria Assessment project}.}


\bibliographystyle{IEEEtran}
\bibliography{bib}

\begin{IEEEbiography}[{\includegraphics[width=1in,height=1.25in,clip,keepaspectratio]{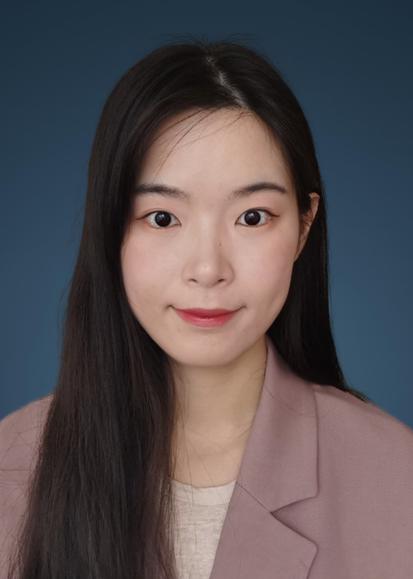}}]{Nan Sun} 
received the B.S. degree (Hons.) and the Ph.D. degree in Information Technology from Deakin University. She is currently a Lecturer in the School of Engineering and Information Technology at the University of New South Wales (UNSW), Canberra, Australia. Before joining UNSW,she was a postdoctoral research fellow at Deakin University. Her current research interests include cybersecurity and social network security.
\end{IEEEbiography}

\begin{IEEEbiography}[{\includegraphics[width=1in,height=1.25in,clip,keepaspectratio]{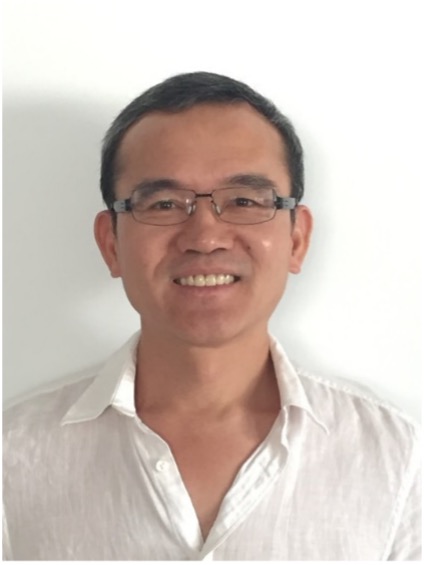}}]{Chang-Tsun Li} received the BSc degree in electrical engineering from National Defence University (NDU), Taiwan, in 1987, the MSc degree in computer science from U.S. Naval Postgraduate School, USA, in 1992, and the PhD degree in computer science from the University of Warwick, UK, in 1998. He was an associate professor of the Department of Electrical Engineering at NDU during 1998-2002 and a visiting professor of the Department of Computer Science at U.S. Naval Postgraduate School in the second half of 2001. He was a professor of the Department of Computer Science at the University of Warwick (UK) until January 2017 and a professor of Charles Sturt University (Australia) from January 2017 to February 2019. He is currently a professor of the School of Information Technology at Deakin University, Australia. His research interests include multimedia forensics and security, biometrics, data mining, machine learning, data analytics, computer vision, image processing, pattern recognition, bioinformatics, and content-based image retrieval. The outcomes of his multimedia forensics research have been translated into award-winning commercial products protected by a series of international patents and have been used by a number of police forces and courts of law around the world. He is currently the Chair of IAPR Computational Forensics Technical Committee, the Associate Editor of IEEE Transactions on Circuits and Systems for Video Technology, EURASIP Journal of Image and Video Processing (JIVP), and IET Biometrics. 
\end{IEEEbiography}
\begin{IEEEbiography}[{\includegraphics[width=1in,height=1.25in,clip,keepaspectratio]{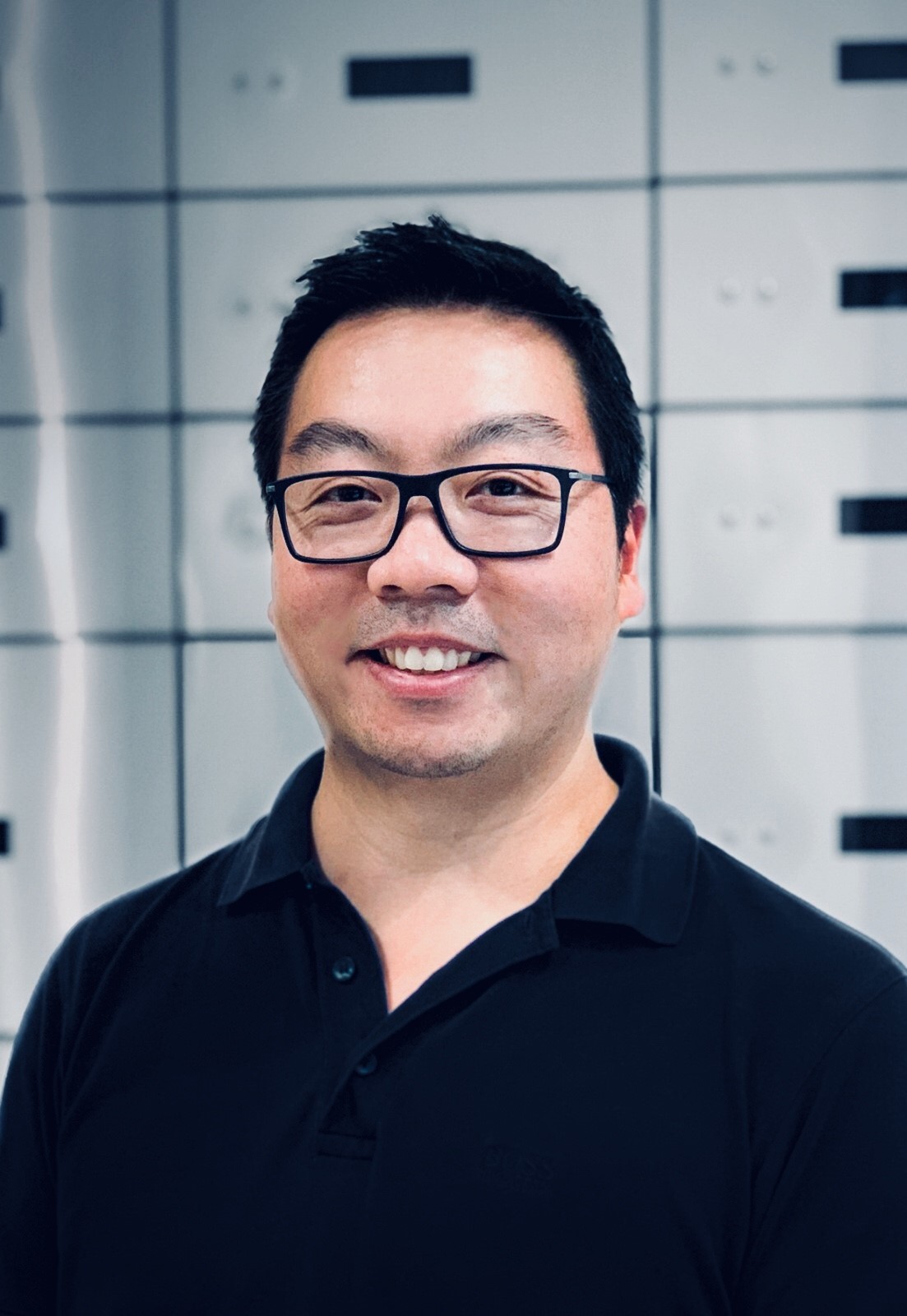}}]{Hin Chan} is the manager of the Australian Information Security evaluation Program (AISEP) that resides within the Australian Cyber Security Centre (ACSC). The AISEP performs Common Criteria (CC) evaluation and certification of ICT security products for Australian Organizations use as well as to set standards to improve the security in ICT products. Within this role, he is the Australian government adviser on all matters related to product assurance and leads the strategic direction of Australia’s international Common Criteria effort. Hin is also an Australian representative at various international CC committees, at ISO JTC1/SC27 working groups and is a member of the Accreditation Advisory Committee (AAC) within Australia’s national accreditation body for testing laboratories, the National Association of testing and Accreditation (NATA).
\end{IEEEbiography}

\begin{IEEEbiography}[{\includegraphics[width=1in,height=1.25in,clip,keepaspectratio]{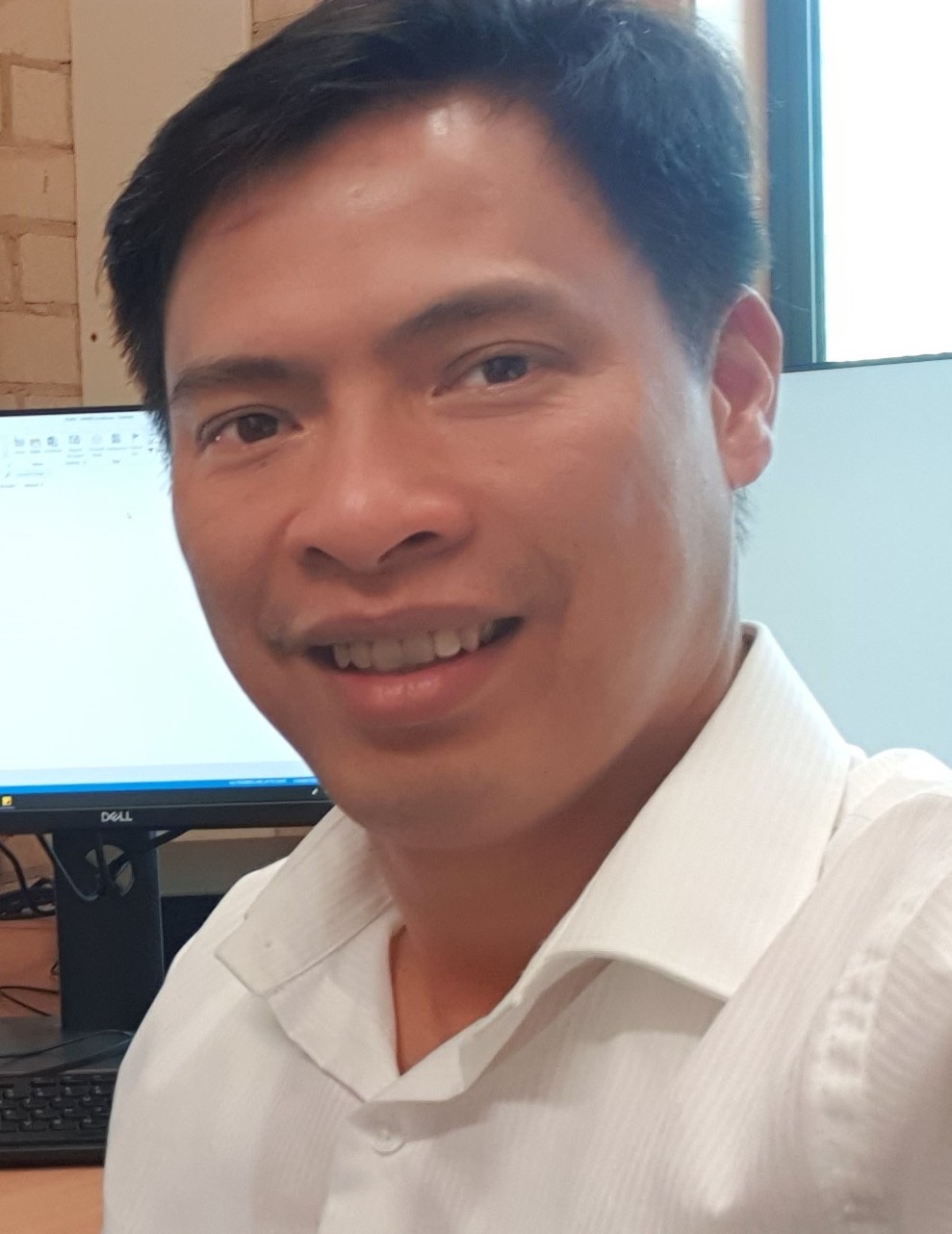}}]{Ba Dung Le} received his Ph.D. in Computer Science from the University of Adelaide, Australia. He worked as a Postdoctoral research fellow in Cyber Security at the Charles Sturt University.  He has led and participated in a number of research projects in Data Clustering, Cyber Threat Detection, and Privacy-Preserving Statistical Aggregation.
\end{IEEEbiography}
\begin{IEEEbiography}[{\includegraphics[width=1in,height=1.25in,clip,keepaspectratio]{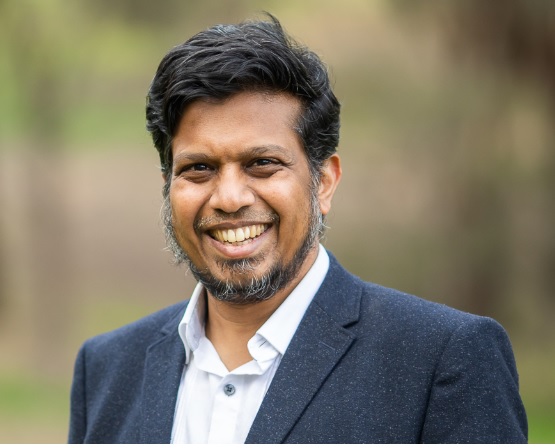}}]{MD ZAHIDUL ISLAM} is a Professor of computer science in the School of Computing, Mathematics, and Engineering, Charles Sturt University, Australia. His main research interests include data mining/machine learning, privacy preserving data mining, applications of data mining/machine learning in real life including cyber security.
\end{IEEEbiography}

\begin{IEEEbiography}[{\includegraphics[width=1in,height=1.25in,clip,keepaspectratio]{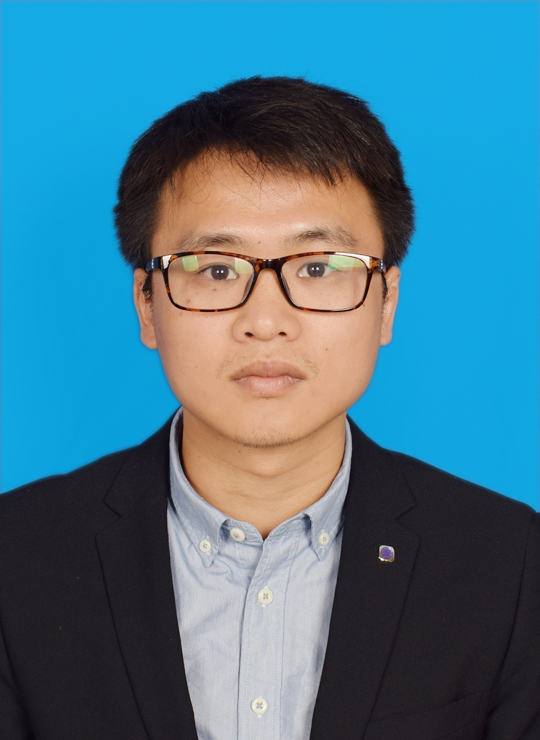}}]{Leo Yu Zhang} (M’17) is currently a Lecturer with the School of Information Technology, Deakin University, VIC, Australia. He received the bachelor’s and master’s degrees in computational mathematics from Xiangtan University, Xiangtan, China, in 2009 and 2012, respectively, and the Ph.D. degree from the City University of Hong Kong, Hong Kong, in 2016. Prior to joining Deakin, he held various research positions with the City University of Hong Kong, the University of Macau, Macau, China, the University of Ferrara, Ferrara, Italy, and the University of Bologna, Bologna, Italy. His current research interests include applied cryptography and AI-related security, and he has published more than 60 refereed journal and conference articles in these fields. 
\end{IEEEbiography}

\begin{IEEEbiography}[{\includegraphics[width=1in,height=1.25in,clip,keepaspectratio]{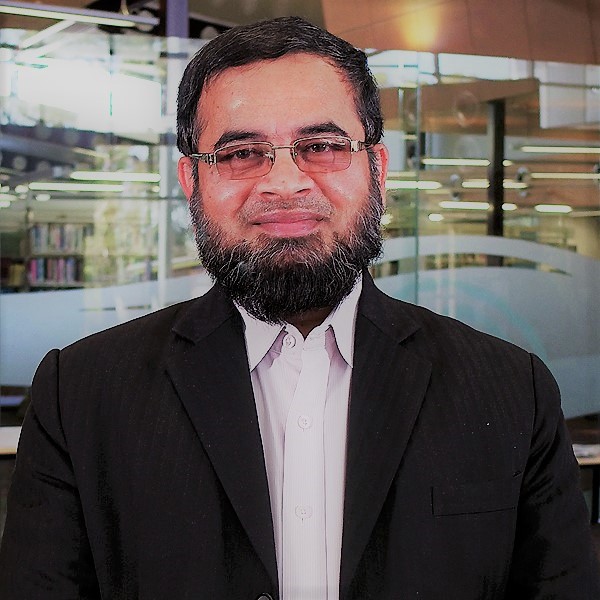}}]{MD Rafiqul Islam} is working as an Associate Professor at the School of Computing, Mathematics and Engineering, Charles Sturt University, Australia. Dr Islam’s main research background in cybersecurity focuses on malware analysis and classification, security in the cloud, privacy in social media, and the dark web.
\end{IEEEbiography}

\begin{IEEEbiography}[{\includegraphics[width=1in,height=1.25in,clip,keepaspectratio]{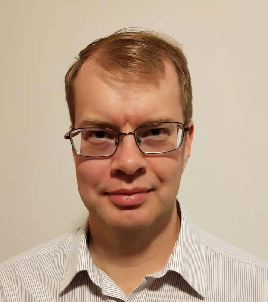}}]{Warren Armstrong} received his PhD from the Australian National University in 2011, and is currently Director of Engineering at QuintessenceLabs, building cyber security products.
\end{IEEEbiography}
\EOD

\end{document}